\documentclass{JHEP3}
\usepackage[utf8]{inputenc}
\usepackage{amstext}
\usepackage{graphicx}
\usepackage{amssymb}
\usepackage{esint}

\newcommand{\OM}{\Omega_m}

\newcommand{\OR}{\Omega_R}
\newcommand{\ORo}{\Omega_R^0}

\newcommand{\rc}{\rho_c}

\newcommand{\OLo}{\Omega_{\Lambda}^0}

\newcommand{\rco}{\rho_{c}^0}
\newcommand{\mincir}{\raise
-3.truept\hbox{\rlap{\hbox{$\sim$}}\raise4.truept\hbox{$<$}\ }}
\newcommand{\magcir}{\raise
-3.truept\hbox{\rlap{\hbox{$\sim$}}\raise4.truept\hbox{$>$}\ }}
\newcommand{\be}{\begin{equation}}
\newcommand{\ee}{\end{equation}}
\newcommand{\rL}{\rho_{\Lambda}}
\newcommand{\rLi}{\rho_{\Lambda}^{i}}
\newcommand{\rLe}{\rho_{\Lambda{\rm eff}}}
\newcommand{\pLe}{p_{\Lambda{\rm eff}}}

\newcommand{\rF}{\rho_{F}}
\newcommand{\pF}{p_{F}}
\newcommand{\GB}{{\cal G}}

\newcommand{\M}{{\cal M}}

\newcommand{\rLo}{\rho_{\Lambda}^0}
\newcommand{\CC}{\Lambda}

\newcommand{\weff}{\omega_{\rm eff}}
\newcommand{\wF}{\omega_{F}}

\newcommand{\newtext}[1]{\text{#1}}

\title{Confronting the relaxation mechanism for a large cosmological constant
with observations}

\author{%
Spyros Basilakos$^{a,b}$, Florian Bauer$^{a,c}$, Joan Solà$^{a,c}$\\
$^a$ High Energy Physics Group, Dept.\ ECM, Univ.\ de Barcelona,
Av.\ Diagonal 647, E-08028 Barcelona,
Catalonia, Spain\\
$^{b}$ Academy of Athens, Research Center for Astronomy and Applied
Mathematics, Soranou Efesiou 4, 11527, Athens, Greece\\
$^{c}$ Institut de Ciències del Cosmos, Univ.\ de Barcelona\\

E-Mails: \email{svasil@academyofathens.gr},
\email{fbauerphysik@eml.cc}, \email{sola@ecm.ub.es}}

\abstract{%
In order to deal with a large cosmological constant a relaxation
mechanism based on modified gravity has been proposed recently. By
virtue of this mechanism the effect of the vacuum energy density of
a given quantum field/string theory (no matter how big is its
initial value in the early universe) can be neutralized dynamically,
i.e.\ without fine tuning, and hence a Big Bang-like evolution of the
cosmos becomes possible. Remarkably, a large class $\{F^n_m\}$ of
models of this kind, namely capable of dynamically adjusting the
vacuum energy irrespective of its value and size, has been
identified. In this paper, we carefully put them to the experimental
test. By performing a joint likelihood analysis we confront these
models with the most recent observational data on type Ia supernovae
(SNIa), the Cosmic Microwave Background (CMB), the Baryonic Acoustic
Oscillations (BAO) and the high redshift data on the expansion rate,
so as to determine which ones are the most favored by observations.
We compare the optimal relaxation models $F^n_m$ found by this
method with the standard or concordance $\CC$CDM model, and find that some
of these models may appear as almost indistinguishable from it.
Interestingly enough, this shows that it is possible to construct
viable solutions to the tough cosmological fine tuning problem with
models that display the same basic phenomenological features as the
concordance model.
}
\keywords{dynamical dark energy, cosmological constant}

\begin{document}

\section{Introduction}

Current observations~\cite{WMAP,SNIa} indicate that the cosmological
constant (CC) $\CC$, or equivalently the vacuum energy density
$\rL=\CC/(8\pi G_N)$, is non-vanishing and positive, and of the
order of $\rLo\sim10^{-47}\,\text{GeV}^{4}$. This value is close to
$70\%$ of the present critical energy density in our universe,
$\rco$, and therefore is very small in Particle Physics units (being
equivalent to having a mass density of a few protons per cubic
meter). In itself, the inclusion of the tiny value~$\rLo$ as a
cosmological term in the gravity action would not be a problem.
However, the modern fundamental physical theories suggest that large
contributions to the classical CC parameter (in particular those
sourced by quantum fluctuations of the matter fields) are induced,
resulting into a huge initial CC value of order $|\rLi|\sim
M_{X}^{4}$ which is fed into the cosmos already in the phase
transitions of the early universe, with~$M_{X}$ in the range between
the electro-weak (EW) scale $\sim 100\,\text{GeV}$ and the Planck
scale~$\sim 10^{19}\,\text{GeV}$. Even if ``only'' the EW scale of
the Standard Model (SM) of Particle Physics would be involved in the
induced CC value, the discrepancy with respect to the observed
value~$\rLo$
entails $55$ orders of magnitude%
\footnote{Let us note that the QCD scale of the strong interactions,
$\Lambda_{\rm QCD}\sim 100~\text{MeV}$, could also trigger an
induced vacuum energy density $\rL\sim \Lambda_{\rm QCD}^4$, which
would then be roughly $43$ orders of magnitude larger than $\rLo$.
We have nevertheless taken the EW value because it is larger and is
considered also as a robust fundamental scale in
the structure of the SM of Particle Physics.}. %
This is of course preposterous. Despite none of these vacuum
contributions is directly supported by experimental evidence yet,
there is hardly a theory beyond the SM that does not come with a
large value of $\CC=8\pi\,G_N\,\rL$ induced by the huge vacuum
energy density $\rL$ predicted by the theory and whose origin may be
both from the zero point vacuum fluctuations of the matter fields as
well as from the spontaneous symmetry breaking of the gauge
theories. In principle, all models with scalar fields (in particular
the SM with its Higgs boson sector) end up with a large vacuum
energy density $\rL$ associated to their potentials. For example,
from the SM Higgs potential we expect $\rL\sim 10^8$ GeV$^4$, which
explains the aforementioned $55$ orders of magnitude discrepancy and
corresponding 55\textsuperscript{th} digit fine tuning CC problem.
The huge hierarchy between the predicted and the observed CC, the
so-called ``old CC problem''~\cite{CCP}, is one of the biggest
mysteries of theoretical physics of all times. To avoid removing the
huge initial value of the vacuum energy density~$\rLi$ by hand with
an extremely fine tuned counterterm\,\footnote{See e.g.\ a detailed
account of the old fine tuning CC problem in Sect.~2, and
Appendix~B, of Ref.\,\cite{Relax-FRG-JCAP}.}  (giving the theory a
very unpleasant appearance), alternative dark energy models have
been suggested. Typically, these models replace the large CC by a
dynamical energy source [see~\cite{DarkEnergy} and references
therein]. This seems reasonable for describing the observed late-time
acceleration, but generally it still requires that the large CC has
been removed somehow. In the end, it means that a fine tuned
counterterm has been tacitly assumed~\cite{CCandDE}.

It is encouraging that time-varying vacuum energy models inspired by
the principles of QFT can be suggested and may hopefully provide an
alternative and more efficient explanation for the dynamical nature
of the vacuum energy\,\cite{JSP11a}. Recently the analysis of some
well motivated models of this kind versus the observations has shown
that their phenomenological status is perfectly reasonable and
promising\,\cite{GSBP11,BPS09a} -- see also the recent works
\cite{Maggiore11}. However, in order to solve the old CC problem in
this context, we need to focus on models that dynamically counteract
the large initial $\rLi$. The idea, in a nutshell, is: 1) to obtain
an effective $\rLe(t)$ (the measured one) which satisfies
$|\rLe(t)|\ll|\rLi|$ at all times $t$ and without fine-tuning; 2) to
insure that $\rLe(t)$ preserves the standard cosmic evolution (i.e.\
the correct sequence of radiation, matter and dark energy dominated
epochs); and finally 3)  to make sure that $\rLe(t)$ is able to
reproduce the measured value of the vacuum energy at present,
$\rLe(t_0)\sim\rLo$, again without fine-tuning. Only in this way we
can guarantee a low-curvature universe similar to ours, namely  $R=
8\pi\,G_N\rLe\sim8\pi\,G_N\rco\sim H_0^2$, with $H_0\sim 10^{-42}$
GeV the current Hubble rate. The question of course is: is that program really possible?

Obviously, in order to follow this road in a successful way it is
necessary to go beyond the common dark energy models and face a
class of scenarios where the large CC is not hidden under the rug,
so to speak, but rather it is permanently  considered as a
fundamental ingredient of the overall theory of the cosmos. In this
work, we analyze a class (probably not unique) of models operating
along this line, namely the CC relaxation mechanism based on
$F(R,\GB)$ modified
gravity~\cite{Relax-FRG-JCAP,Relax-FRG-PLB,Relax-FRG-Astro}. These
models are intimately related with the family of the so-called $\Lambda$XCDM
models of the cosmic evolution\,\cite{LXCDM} (in which $X$ is
generally \textit{not} a scalar field, but an effective quantity in
the equations of motion which derives from the complete structure of
the effective action). They might also be connected with mechanisms
based on matter with an inhomogeneous equation of
state~\cite{Relax-inhEOS}. A first modified gravity approach
implementing the relaxation mechanism was presented in
\,\cite{Relax-LXCDM-1}. Furthermore, a related model in the
alternative Palatini formalism has been studied~\cite{Palatini-CC},
too.

As stated, the class of models considered here is probably just a
subset of a larger class of theories that can produce dynamical
relaxation of the vacuum energy.  Our work should ultimately
be viewed as aiming at illustrative purposes only, i.e.\ as
providing a proof of existence that one can construct explicitly a
relaxation mechanism, if only in a moderately realistic
form\,\footnote{Remember that in the past the dynamical adjustment
of the CC was attempted through scalar field
models\,\cite{OldScalar}, but later on a general no-go theorem was
formulated against them\,\cite{CCP}. Fortunately, this theorem can
be circumvented by the relaxation mechanism, see
\cite{Relax-FRG-JCAP} for details.}. We cannot exclude that more
sophisticated and efficient mechanisms can be eventually discovered
which are much more realistic, but at the moment the $F(R,\GB)$
models serve quite well our illustrative purposes. They constitute a
potentially important step in the long fighting of Theoretical
Physics against the tough CC problem, especially in regard to the
appalling fine tuning conundrum inherent to it. Additional work
recently addressing the CC problem from various perspectives can be
found e.g.\ in Refs.~\cite{CCP-Solutions}.

In contrast to late-time gravity modifications, in the relaxation mechanism an effective energy density~$\rho_{F}(t)$ is permanently induced by the $F(R,\GB)$ modification to the standard gravity action, which counteracts dynamically (at any time of the postinflationary cosmic history) the
effect of the large vacuum energy density~$\rLi$ and provides the
net value $\rLe=\rLi+\rF$  (presumably close to the
observationally measured one $\rLo$). As a result the universe has
an expansion history similar, but not identical, to the
concordance~$\Lambda$CDM model. Therefore, it is important to
investigate how the relaxation models perform with respect to
observational constraints coming from  the Cosmic Microwave
Background (CMB)\,\cite{WMAP}, type Ia supernovae
(SNIa)\,\cite{SNIa}, the Baryonic Acoustic Oscillations
(BAO)\,\cite{Eis05} and the high redshift data on the expansion
rate\,\cite{Hubble-Observ}.

The paper is organized as follows: in Sect.~\ref{sec:CCRelax-Model}
we discuss different aspects of the working principle of the CC relaxation mechanism in modified gravity. In Sect.~\ref{sec:Likelihood} we present the numerical tools that we use to perform the statistical
analysis, and in Sect.~\ref{sec:BestModels} we apply them to
determine the optimal relaxation model candidates with a further
insight in the details of the relaxation mechanism. In
Sect.~\ref{sec:ComparingLCDM} we present predictions for the
redshift evolution of the deceleration parameter and the effective
equation of state, and compare with the $\CC$CDM. In the last
section we deliver our conclusions. Finally, in an appendix we
provide some useful formulae discussed in the text.

\section{Vacuum relaxation in $F(R,\GB)$ modified gravity}\label{sec:CCRelax-Model}

The general form of the complete effective action of the
cosmological model in terms of the Ricci scalar~$R$ and the
Gauss-Bonnet invariant~$\GB$ is given by
\begin{equation}
\mathcal{S}=\int d^{4}x\,\sqrt{|g|}\left[\frac{1}{16\pi
G_{N}}R-\rLi-F(R,\GB)+\mathcal{L}_{\phi}\right],\label{eq:CC-Relax-action}
\end{equation}
where ${\cal L}_{\phi}$ denotes the Lagrangian of the matter fields.
This action represents the Einstein-Hilbert action extended by the
term~$F(R,\GB)$ defining the modification of gravity. Moreover, we
include all vacuum energy contributions from the matter sector in
the large CC term~$\rLi$.

From the above action the modified Einstein equations are derived by
the variational principle $\delta\mathcal{S}/\delta g^{ab}=0$. They
read
\begin{equation}
G_{\,\, b}^{a}=-8\pi G_{N}\left[\rLi\,\delta_{\,\, b}^{a}+2E_{\,\,
b}^{a}+T_{\,\, b}^{a}\right],\label{eq:Mod-Einstein-Eqs}
\end{equation}
with the Einstein tensor~$G_{\,\, b}^{a}=R_{\,\,
b}^{a}-\frac{1}{2}\delta_{\,\, b}^{a}\, R$, the cosmological term
and the energy-momentum tensor~$T_{\,\, b}^{a}$ of matter emerging
from~$\mathcal{L}_{\phi}$.

We describe the cosmological background by the spatially flat FLRW
line element $ds^{2}=dt^{2}-a^{2}(t)d\vec{x}^{\,2}$ with the scale
factor~$a(t)$ and the cosmological time~$t$. Accordingly, the
curvature invariants~$R=6H^{2}(1-q)$ and~$\GB=-24H^{4}q$ can be
expressed in terms of the Hubble expansion rate~$H=\dot{a}/a$ and
the deceleration parameter~$q=-\ddot{a}a/\dot{a}^{2}$. The tensor
components in (\ref{eq:Mod-Einstein-Eqs}) are given by
$G_{\,\,0}^{0}=-3H^{2}$, $G_{\,\, j}^{i}=-\delta_{\,\,
j}^{i}(2\dot{H}+3H^{2})$ and
\begin{eqnarray}
E_{\,\,0}^{0} & = & \frac{1}{2}\left[F-6(\dot{H}+H^{2})F^{R}+6H\dot{F}^{R}-24H^{2}(\dot{H}+H^{2})F^{\GB}+24H^{3}\dot{F}^{\GB}\right]\label{eq:E00}\\
E_{\,\, j}^{i} & = & \frac{1}{2}\,\delta_{\,\, j}^{i}\left[F-2(\dot{H}+3H^{2})F^{R}+4H\dot{F}^{R}+2\ddot{F}^{R}\right.\nonumber \\
 &  & \left.-24H^{2}(\dot{H}+H^{2})F^{\GB}+16H(\dot{H}+H^{2})\dot{F}^{\GB}+8H^{2}\ddot{F}^{\GB}\right],\label{eq:Eij}
\end{eqnarray}
where $F^{Y}\equiv\partial F/\partial Y$ stand for the partial
derivatives of $F$ with respect to~$Y=R,\GB$. The $F$-term induces
the effective energy density~$\rho_{F}=2E_{\,\,0}^{0}$ and
pressure~$p_{F}=-\frac{2}{3}E_{i}^{\,\,i}$, implying that the whole
effective dark energy density and pressure consist of two parts
each:
\begin{equation}
\rLe(t)=\rLi+\rho_{F}(t),\,\,\,\,\,\,\,\,\,
\pLe(t)=-\rLi+p_{F}(t).\label{eq:rhoLambdaEff}
\end{equation}
As announced, they follow the structure of the $\Lambda$XCDM
model\,\cite{LXCDM}, as both are the sum of a cosmological term
energy density (resp. pressure) and an extra contribution whose
structure derives from the presence of new terms in the effective
action. The local energy density $\rF$ constitutes the ``induced DE
density'' for this model. It is indeed induced by the new term of
the modified gravity action (\ref{eq:CC-Relax-action}). However, the
only measurable DE density in our model is the effective vacuum
energy $\rLe$ in (\ref{eq:rhoLambdaEff}), i.e.\ the sum of the
initial (arbitrarily large) vacuum energy density $\rLi$ and the
induced DE density $\rF$. None of the latter, though, is individually
measurable. Let us note that the induced DE density is covariantly
self-conserved, i.e.\ independently of matter (which is also
conserved). Indeed, the Bianchi identity on the FLRW background
leads to the local covariant conservation laws
\begin{equation}
\dot{\rho_n}+3H(\rho_n+p_n)=0\,,\label{eq:Bianchi}\,\end{equation}
which are valid for all the individual components ($n=$ radiation,
matter and vacuum) with energy density~$\rho_n$ and pressure~$p_n$.
This leads immediately to the usual expressions
$\rho_{m}=\rho_{m}^{0}a^{-3}$ and $\rho_{r}=\rho_{r}^{0}a^{-4}$ for
cold matter and radiation respectively. As for the effective vacuum
energy density $\rLe$, since $\rLi$ is a true CC term and satisfies
$p_{\CC}^i=-\rho_{\CC}^i$, it follows that the $F$-term density is
self-conserved:
\begin{equation}\label{localCCF}
\dot{\rho}_{F}+3H(\rF+\pF)=\dot{\rho}_{F}+3H\,\left[1+\wF(t)\right]\rF=0\,,
\end{equation}
where $\wF=\wF(t)$ is the corresponding (non-trivial) effective
equation of state~(EoS) of the $F$-term. We shall come back to it in more detail in Sect.~\ref{sec:ComparingLCDM}.

The cosmological evolution is therefore completely determined by the
generalized Friedmann equation of our model,
\begin{equation}
\frac{3H^{2}}{8\pi
G}=\rho_{m}+\rLe=\rho_{m}+\rLi+\rho_{F},\label{eq:Friedmann1}
\end{equation}
where~$\rho_{m}\propto a^{-3}$ denotes the energy density of the
(self-conserved) matter, which is mostly dust-like matter for the
purpose of this paper (as we focus mainly on the matter dominated
and DE epochs).

Next, we consider the CC relaxation mechanism from
Refs.~\cite{Relax-FRG-JCAP,Relax-FRG-PLB}, in which $F$ in the
action~(\ref{eq:CC-Relax-action}) is picked within the class of
functions $F^n_{m}=F^n_{m}(R,\GB)$ of the form
\begin{equation}
F^n_m=\beta\frac{R^{n}}{\left(B(R,\GB)\right)^{m}}\equiv\beta\frac{R^{n}}{\left[\frac{2}{3}R^{2}+\frac{1}{2}\GB+(y\,
R)^{3}\right]^{m}}\,.\label{eq:Fmn}
\end{equation}
Here $\beta$ and~$y$ are two parameters of the model, and $n\geq0$
and~$m>0$ (usually taken to be
integers) are two numbers characterizing a large class of functions
that can realize the relaxation mechanism. To understand how the mechanism works in modified
gravity, let us express the denominator of (\ref{eq:Fmn}) in terms
of~$H$ and~$q$,
\begin{equation}
B(H,q)=24H^{4}(q-\frac{1}{2})(q-2)+\left[6\, y\,
H^{2}(1-q)\right]^{3}.\label{eq:B}
\end{equation}
Taking into account that each derivative of~$F$ in
Eqs.~(\ref{eq:E00}) and~(\ref{eq:Eij}) increases the power of the
denominator function~$B$, it is easy to see that the induced energy
density~$\rho_{F}$ and pressure~$p_{F}$ have at least one positive
power of~$B$ in their denominators. Schematically, one can show that
it takes on the form
\begin{equation}\label{eq:rhofi}
\rF(t)=\frac{f_1(t)}{B^{m}}+\frac{f_2(t)}{B^{m+1}}+\frac{f_3(t)}{B^{m+2}}\,,
\end{equation}
with $f_i$ being time-dependent functions not containing the
factor~$B$. Notice that we need $n<2m$ in order that $\rF$ can
increase with the expansion, i.e.\ when $H$ is decreasing. In this
case, the gravity modification~$F$ in the action
(\ref{eq:CC-Relax-action}) will be able to compensate the large
initial vacuum energy density~$\rLi$ in the field equations
(\ref{eq:Mod-Einstein-Eqs}), whatever it be its value and size, by
$\rho_{F}=2\,E^0_{\,\,0}$ becoming large in a dynamical manner. It
means that $\rho_{F}$ can take the necessary value to compensate for
the big initial $\rLi$, this being achieved automatically
because~$B$ becomes small during the radiation dominated regime
($q\to 1$), then also in the matter dominated era
($q\to\frac{1}{2}$), and finally again in the current universe and
in the asymptotic future, where~$H$ becomes very tiny.

\newtext{A few additional observations are in order.} The fact that we search our relaxation functionals among those of the form $F(R,\GB)$ is because they are better behaved. Indeed, let us recall that general modified gravity
models of the form $F(R,S,T)$, with $R=g^{ab}R_{ab}$,
$S=R_{ab}R^{ab}$ and $T=R_{abcd}R^{abcd}$, are generally problematic
as far as ghosts and other instabilities are concerned. This is because these theories introduce new degrees of freedom, which potentially lead to instabilities not present in general relativity (GR). They may e.g. suffer from the Ostrogradski instability, i.e. they may involve vacuum states of negative energy. In general they contain gravitational ghosts and can lead to various types of singularities. These issues are discussed e.g. in \cite{Carroll04} and are reviewed in \cite{SotiriouFaraoni08}. Fortunately, some of these problems can be avoided by specializing to $F(R,\GB)$
functionals involving only the Ricci scalar and the Gau\ss-Bonnet
invariant\,\cite{Bamba09}. The reason is that all functionals of the form
$F(R,T-4S)$ are ghost free\,\cite{Comelli05}, a property which is shared by our functionals because $\GB=R^{2}-4S+T$.

\newtext{Finally}, the $F(R,\GB)$  theories are thought to have a more reasonable behavior in the solar system limit~\cite{Comelli05,Nojiri:2005jg}. A detailed study of the astrophysical consequences of our model was presented in \cite{Relax-FRG-Astro}. It shows that a huge cosmological constant can also be relaxed in astrophysical systems such as the solar system
and the galactic domain. These studies were performed by solving the
field equations in the Schwarzschild-de Sitter metric in the
presence of an additional $1/R$ term. The latter operates the
relaxation of the huge CC at astrophysical scales in a similar way as the $\sim 1/B$ term in equation (\ref{eq:Fmn}) does it in the cosmological domain. One can show that none of these terms has any significant influence in the region of dominance of the other. Therefore a huge CC can be reduced both in the local astrophysical scales and in the
cosmological one in the large. The solution in the Schwarzschild-de Sitter metric merges asymptotically with the cosmological solution, where we
recover the original framework presented in \cite{Relax-FRG-JCAP}.
Furthermore, it was shown in \cite{Relax-FRG-Astro} that there are no additional (long range) macroscopic forces that could correct in a measurable way the standard Newton's law. This is a consequence of the dynamical relaxation mechanism and is in contrast to ordinary modified gravity models. The upshot is that the implementation of the relaxation mechanism should not be in conflict with local gravitational experiments. At the same time, we expect small deviations from GR only at scales of a few hundred kpc at least, which is in accordance with other authors\,\cite{NavarroAcoleyen05}. As this scale is much larger than the scale where star systems can be tested at the level of GR, the physics of these local astrophysical objects should not be affected.

\subsection{Physical scales for the relaxation mechanism}\label{sect:PhysicalScales}

Remarkably, in order to insure that~$|\rLi+\rho_{F}|\ll|\rLi|$ holds
at all times, we need not fine tune the value of any parameter of
the model. It is only necessary that the parameter~$\beta$ in the
class of invariants (\ref{eq:Fmn}) has the right order of magnitude
and sign. This requirement has nothing to do with fine tuning, as
fine tuning means to adjust by hand the value of the parameter
$\beta$ to some absurd precision such that both (big) terms $\rLi$
and $\rF$ almost cancel each other in (\ref{eq:rhoLambdaEff}). This
is not necessary at all for our mechanism to work because it is
dynamical, meaning that it is the evolution itself of the universe
through the generalized Friedmann equation~(\ref{eq:Friedmann1})
what drives the $F$-term density $\rho_{F}$ towards almost canceling
the huge initial $\rLi$ in each relevant stage of the cosmic
evolution (i.e.\ in the radiation dominated, matter
dominated and dark energy dominated epochs). In this way we can
achieve the natural relation
\begin{equation}
\frac{\rLe}{\rLi}=
\frac{\rLi+\rho_{F}(H_{*})}{\rLi}=\frac{\rco-\rho_{m}^0}{\rLi}=\mathcal{O}\left(\frac{\rco}{\rLi}\right)\ll
1,\label{eq:Friedmann-Relax}
\end{equation}
for some value of $H_{*}$ sufficiently close to $H_0$. The smallness
of the observed~$\rLe$ as compared to the huge initial $\rLi$ thus
follows from the right-hand side of (\ref{eq:Friedmann-Relax}) being
suppressed by the ratio of the present critical density
$\rco=3H_0^2/(8\pi G)$ as compared to its initial value in the early
universe, which was of course of order $\rLi$. By working out the
explicit structure of (\ref{eq:rhofi}) from (\ref{eq:E00}), and
using the fact that $\dot{H}=-(q+1)H^2$, we can easily convince
ourselves that all terms end up roughly in the form $\rF(H)\sim
\beta\,H^{2n-4m}$. Although we are omitting here other terms, we
adopt provisionally that expression for the sake of simplicity (see
the next section for more details). Within this simplified setup, it
follows that the value of $H_{*}$ that solves equation
(\ref{eq:Friedmann-Relax}) -- which, as we said, should be close
enough to the current value of $H$ -- is approximately given by
\begin{equation}\label{eq:Hstar}
H^2_{*}\simeq \left|\frac{\beta}{\rLi}\right|^{1/(2m-n)}\,.
\end{equation}
We see that for $n<2m$ this mechanism provides also a natural
explanation for the current value of $H$ being so small: the reason
simply being that $\rLi$ is very large! Moreover, we can make
$H_{*}$ of order of the measured $H_0$ provided $\beta$ has the
right order of magnitude (without operating any fine tuning). Notice
that since the power mass dimension of  $\beta$ for the $F^n_m$
models is
\begin{equation}\label{eq:Mpower}
|\beta|=\M^{4-2n+4m}\,,
\end{equation}
and $\rLi\sim M_X^4$, it follows that $H_{*}\equiv H_0$ is related
to $\M$ through
\begin{equation}\label{eq:MH0}
H_0\simeq \M\,\left(\frac{\M}{M_X}\right)^{2/(2m-n)}\,.
\end{equation}
For the simplest $F^0_1$ model this implies $\M\sim\sqrt{M_X H_0}$,
and therefore if $M_X$ is near $M_P$ we obtain $\M$ around the
characteristic meV scale associated to the current CC density:
$m_{\CC}\equiv \left(\rLo\right)^{1/4}\sim 10^{-3}$ eV. On the other
hand, for the models $F^1_1$ and $F^3_2$ we find in both cases
$\M\sim \left(M_X^2\,H_{0}\right)^{1/3}$, which implies $\M\sim
0.1-100\,\text{MeV}$ for $M_X$ in the ballpark of the GUT scale
($\sim 10^{16}$ GeV) up to the Planck mass $M_P\sim 10^{19}$ GeV.
This is certainly a possible mass scale in the SM of Particle
Physics.

Notice that equations (\ref{eq:Hstar}) and (\ref{eq:MH0}) have been
derived under the assumption $n\neq 2m$. Therefore, the conclusion
that the current value of $H$ is small because $\rLi$ is large
cannot be inferred from equation (\ref{eq:Hstar}) if $n=2m$. Still,
we shall see in the next next section that a more accurate treatment
of the $n=2m$  models leads once more to the conclusion that the current $H$ is small. What else can be learnt from equation (\ref{eq:MH0})? Notice
that it can be rewritten as
\begin{equation}\label{eq:MH0bis}
2m-n\simeq
2\,\frac{\ln\left(M_X/\M\right)}{\ln\left(\M/H_0\right)}\simeq
\frac{2(s-r)}{r+42}\,,
\end{equation}
with $M_X\equiv 10^s$ GeV, $\M\equiv 10^r$ GeV and $H_0\sim
10^{-42}$ GeV (for some integers $r$ and $s$). Now, being
$M_X\leqslant M_P\sim 10^{19}$ GeV the GUT scale, we must have
$r\leqslant s\leqslant 19$. For $r\geqslant 0$, it follows from
(\ref{eq:MH0bis}) that $0\leqslant 2m-n\lesssim 1$. But let us keep
in mind that we could have $r<0$. Since, however, the scale $\M$
should not be too tiny\,\footnote{If $\M$ could be very small, we
would stumble upon the same tiny mass problem that afflicts e.g.\
all quintessence-like models\,\cite{DarkEnergy}, where mass scales
as small as $H_0\sim 10^{-33}$ eV are a common place. This is one, but certainly not
the only one, of the most serious drawbacks of these models, another
one being of course the need of extreme fine-tuning. In our case,
however, we aim precisely at avoiding both of these severe
problems.}, it is natural to assume that it is not much smaller than
the typical meV scale associated to the CC density, i.e.\ we must
have $\M\gtrsim m_{\CC}\sim$ meV$=10^{-12}$ GeV, which enforces
$r\gtrsim -12$. Thus e.g.\ in the extreme case $s=19$ and $r=-12$,
we have $2m-n\lesssim 3$. The upshot of this (order of magnitude)
consideration is that, with $r$ bounded in the approximate interval
$-12\lesssim r\leqslant s$, we should expect in general that the
following combined ``natural relaxation condition'' is fulfilled:
\begin{equation}\label{eq:NatRelax}
0\leqslant 2m-n<{\cal O}(1)\,,
\end{equation}
in which both naturalness of the mass scales and relaxation of the
vacuum energy are insured. In other words, the above argument
suggests that if $n$ and $m$ are taken as integers, we cannot assume
large values for them (unless $n=2m$) because the relation
(\ref{eq:NatRelax}) could not be fulfilled. This is of course a
welcome feature because it suggests that only the canonical cases
are natural candidates, namely $(n,m)=(0,1),(1,1),(2,1),(3,2)$ and
$(2,2)$ at most -- the last case of this series being the one which
approaches the most to the upper bound imposed by the relation
(\ref{eq:NatRelax}), but still the difference is of ${\cal O}(1)$.
The next model in the list, $(n,m)=(2,3)$, tenses up a bit too much
perhaps that bound, so for definiteness we stop the number of
candidate models to the first five in the list. Of course the models
with $n=2m$ for arbitrary $n$ would do, but again it is natural to
focus on only the canonical representative $(n,m)=(2,1)$ of this
class. In general the class $n=2m$ is special because, as we can see
from equation (\ref{eq:MH0bis}), it requires $\M=M_X$. Therefore, if
$M_X\sim 10^{16}$ GeV is a typical large GUT scale, then $\M$ must
coincide with it. We may think of the $n=2m$ models as the class of
``no-scale'' relaxation models inasmuch as they do not introduce any
other new scale beyond the one associated to the initial vacuum
energy itself, $M_X=\left(\rLi\right)^{1/4}$. Also interesting about
the ``no-scale models'' is to note that if there is no GUT scale
above the SM of Particle Physics, i.e.\ if $M_X$ turns out to be just
the electroweak scale $M_W={\cal O}(100)$ GeV, then the relaxation
mass parameter $\M$ will naturally take on the order of magnitude
value $M_W$, characteristic of the electroweak gauge boson masses,
for all the relaxation models $n=2m$. Remarkably, it is the only
situation where we can accommodate the electroweak scale $M_W$ in
the relaxation mechanism.

To summarize, even though we have in principle five canonical
candidate models satisfying the natural relaxation condition
(\ref{eq:NatRelax}), on the whole they involve only two new physical
scales for the dynamical adjustment of the cosmological constant, to
wit: the two models $F^0_1\sim F^2_2$ are both linked to the
same mass scale $\M_1$ close to the one inherent to the
current CC density, $\M_1\sim m_{\CC}\sim 10^{-3}$ eV, whereas the two models
$F^1_1\sim F^3_2$ are both associated to the scale $\M_2\sim
0.1-100$ MeV, which is a typical Particle Physics scale within the
SM of electroweak and strong interactions. Thus both scales quite natural ones.  On the other hand the fifth model $F^2_1$
(and, for that matter, the entire $F^{2m}_m$ class) is a
``no-scale'' model which is able to operate the relaxation mechanism
by using the very same scale as the one associated to $\rLi\sim
M_X^4$, irrespective of $M_X$ being a huge GUT scale or just the
electroweak scale of the SM.

As for the dimensional parameter $y$ in (\ref{eq:B}), let us note
that it is basically irrelevant for the present discussion. In
Ref.\,\cite{Relax-FRG-JCAP}, this parameter played a role only in
the radiation dominated epoch and did not suffer any fine tuning
either; it just had to take a value within order of magnitude. In
practice the whole structure $(y\,R)^3$ of the last term in equation
(\ref{eq:B}) was motivated by considerations related to getting a
smooth transition from the radiation to the matter dominated epochs.
Since, however, we are now comparing the class of models
(\ref{eq:Fmn}) with the observations, all the relevant data to which
we can have some access belong to the matter dominated epoch until
the present time (apart from a correction from the radiation term in the case of the CMB data, which has been duly taken into account, see Sect. \ref{sec:Likelihood}). We have indeed confirmed numerically that the last term on the \textit{r.h.s.} of equation (\ref{eq:B}) is unimportant for the present analysis.

\subsection{A closer look to the general structure of the induced DE density $\rF$}\label{sect:moreFnm}

After we have presented the simplest version of the relaxation
mechanism, it seems appropriate to discuss a bit further the general
behavior of the $F$-density $\rF$ in order to better understand the
relaxation mechanism in the different cases. Following the
considerations exposed at the end of the last section, hereafter we
will set~$y=0$ in equation (\ref{eq:B}). We start by considering the
the general structure of the $F$-density $\rF=2E_{\,\,0}^{0}$ for an
arbitrary $F^n_m$ model, which as we know can be cast as in equation
(\ref{eq:rhofi}). However, by dimensional analysis and the explicit
structure of (\ref{eq:E00}) and (\ref{eq:B}), it is not difficult to
convince oneself that it can eventually be brought into the more
specific form:
\begin{equation}\label{rFgeneral}
{\rF(H,q,\dot{q}/H)}=\frac{\beta}{H^{4m-2n}}\left[f(q)+g(q)\,\frac{\dot{q}}{H}\right]\,.
\end{equation}
Here $f(q)$ and $g(q)$ are functions of $q$ which are different for
different $F^n_m$ models, but in all cases they are rational
functions of $q$ with a (multiple) pole at $q=1/2$, specifically
$f(q)\propto 1/b^k$, $g(q)\propto1/b^{k+1}$, where for convenience
we have defined the following expression that appears in the
calculation:
\begin{equation}\label{eq:b}
b:=2(q-\frac{1}{2})(q-2)\,.
\end{equation}
For specific realizations of equation (\ref{rFgeneral}), see
Appendix~\ref{sect:rFmn}. Let us note that because of the pole at
$q=1/2$ all these models have an unstable fixed point in the matter
dominated epoch, which is responsible for the universe to approach
$q\to 1/2$ for a long while during that epoch and at the same time
enhances dynamically the value of $\rF$. As a result the huge value
of the vacuum energy can be relaxed from $\rLi$ to the effective
tiny $|\rLe|\ll|\rLi|$, which can be identified with~$\rLe=\rLo$ at
the present time. When that pole is left behind during the cosmic
evolution, the relaxation mechanism can still work effectively
provided $H$ tends to a very small value in the present universe.
Notice that there are two terms in equation (\ref{rFgeneral}) that
cooperate to fulfill this end: one of them is the overall $\sim
1/H^{4m-2n}$ factor, which contributes to the relaxation mechanism
provided $n$ is strictly smaller than $2m$, as indeed required by
the relation (\ref{eq:NatRelax}); and the other is the term
$\dot{q}/H$ in the parentheses. For all the models such that $n<2m$,
the first term suffices and in this way we recover the result
(\ref{eq:Hstar}) which we had sketched in the simplified exposition
of Sect.~\ref{sect:PhysicalScales}.

However, if $n=2m$ the simplified argument of Sect.
\ref{sect:PhysicalScales} cannot be applied. The class of these
``no-scale'' functionals is of the form $F^{2m}_m\sim M_X^4
R^{2m}/B^m$, where the relevant parameter is $M_X$, the same one as
that associated to $\rLi\sim M_X^4$. As previously noticed, this is
the only case where the scale $\M$ defined in (\ref{eq:Mpower})
could be identical to the initial vacuum energy \newtext{scale}
$M_X\equiv\left(\rLi\right)^{1/4}\geq 10^{16}$ GeV of the GUT at the early
universe, which is a very interesting possibility in that here the
mechanism for canceling the large vacuum energy density of the early
universe would be naturally dealt with by the very same scale
pertaining to the GUT transition at that time. For the $n=2m$
models, the factor $\dot{q}/H$ inside the parenthesis in equation
(\ref{rFgeneral}) takes its turn in the relaxation mechanism. Noting
that now equation~(\ref{rFgeneral}) simplifies to
\begin{equation}\label{rFmeqn}
\rF(q,\dot{q}/H)=\beta\,\left[f(q)+g(q)\,\frac{\dot{q}}{H}\right]\ \
\ \ \ \ \ (n=2m)\,,
\end{equation}
it follows that the relaxation condition (\ref{eq:Friedmann-Relax})
for the present universe enforces $H_{*}$ to take the value
\begin{equation}\label{Hneqm}
H_{*}=\left|\frac{\beta\,g(q)\,\dot{q}}{\beta\,f(q)+\rLi}\right|\sim\left|\frac{\beta}{\rLi}\,g(q)\,\dot{q}\right|\sim
\left|g(q)\dot{q}\right|\sim \left|\dot{q}\right|\,.
\end{equation}
In this equation we have used the fact that, for the $n=2m$ models,
we have $\beta=M_X^4$, which is of the same order of magnitude as
$\rLi\sim M_X^4$ and therefore the two factors cancel out
approximately. Finally, $g(q)$ is a dimensionless function,
basically a number which can be taken of order $1$ in this kind of
consideration. Therefore, we conclude that the predicted value of
the current expansion rate for these models is $H_{*}\sim
|\dot{q}|$, and therefore it can naturally be of order $H_0$. So,
again, the relaxation mechanism works and requires a very small
value for the present Hubble rate.

\section{Likelihood analysis from CMB, SNIa, BAO and $H(z)$ data}\label{sec:Likelihood}

In the following we briefly present some details of the
statistical method and on the observational samples and data
statistical analysis that will be adopted to constrain the
relaxation models presented in the previous section. First of all,
we use the {\em Constitution} set of 397 type Ia supernovae of
Hicken et al.~\cite{Hic09}. In order to avoid possible problems
related with the local bulk flow, we use a subsample of 366 SNIa,
excluding those with $z<0.02$. The corresponding $\chi^{2}_{\rm
SNIa}$ function, to be minimized, is:
\begin{equation}\label{eq:xi2SNIa}
\label{chi22} \chi^{2}_{\rm SNIa}({\bf p})=\sum_{i=1}^{366} \left[
\frac{ {\cal \mu}_{\rm th} (a_{i},{\bf p})-{\cal \mu}_{\rm
obs}(a_{i}) } {\sigma_{i}} \right]^{2} \;,
\end{equation}
where $a_{i}=(1+z_{i})^{-1}$ is the observed scale factor of the
Universe for each data point and $z_{i}$ the corresponding
(measured) redshift. The fitted quantity ${\cal \mu}$ is the
distance modulus, defined as ${\cal \mu}=m-M=5\log{d_{L}}+25$, in
which $d_{L}(a,{\bf p})$ is the luminosity distance:
\begin{equation}\label{eq:LumDist}
d_{L}(a,{\bf p})=\frac{c}{a} \int_{a}^{1} \frac{{\rm
d}a'}{a'^{2}H(a')}={c}{(1+z)} \int_{0}^{z} \frac{{\rm
d}z'}{H(z')}\;,
\end{equation}
with $c$ the speed of light and ${\bf
p}=(\Omega_{m},q_{0},{\dot{q}_{0}}/{H_{0}})$ a vector containing the
cosmological parameters of our model that we wish to fit for. In
equation (\ref{eq:xi2SNIa}), the theoretically calculated distance
modulus $\mu_{\rm th}$ for each point follows from using
(\ref{eq:LumDist}), in which the Hubble function is given by the
generalized Friedmann's equation (\ref{eq:Friedmann1}). Finally,
$\mu_{\rm obs}(a_{i})$ and $\sigma_i$ stand for the measured
distance modulus and the corresponding $1\sigma$ uncertainty for
each SNIa data point, respectively. The previous formula
(\ref{eq:LumDist}) for the luminosity distance applies only for
spatially flat universes, which we are assuming throughout.


On the other hand, a very interesting geometrical probe of dark
energy is provided by the measures of $H(z)$~\cite{Hubble-Observ}
from the differential ages of passively evolving galaxies [hereafter
$H(z)$ data]\,\footnote{For the current value of the Hubble
constant, we use $H_{0}\equiv 100\,h=73\,\text{km/s/Mpc}$~\cite{HDEmodels}.}.
This sample of galaxies contains 11 entries spanning a redshift
range of $0\le z<2$, and the corresponding $\chi^{2}_{\rm H}$
function can be written as:
\begin{equation}
\chi^{2}_{\rm H}=\sum_{i=1}^{11} \left[ \frac{ H_{\rm th}(z_{i},{\bf
p})-H_{\rm obs}(z_{i})} {\sigma_{i}} \right]^{2} \;\;.
\end{equation}
Note that the latter set of data became recently interesting and
competitive for constraining dark energy, see e.g.~\cite{HDEmodels}.

In addition to the SNIa and $H(z)$ data, we also consider the BAO
scale produced in the last scattering surface by the competition
between the pressure of the coupled baryon-photon fluid and gravity.
The resulting acoustic waves leave (in the course of the evolution)
an overdensity signature at certain length scales of the matter
distribution. Evidence of this excess has been found in the
clustering properties of the SDSS galaxies
(see~\cite{Eis05},~\cite{Perc10}) and it provides a ``standard
ruler'' that we can employ to constrain dark energy models. In this
work we use the measurement derived by Eisenstein et
al.~\cite{Eis05}. In particular, we utilize the following estimator
\begin{equation}
A({\bf p})=
=\frac{\sqrt{\Omega_{m}^0}}{[z^{2}_{s}E(z_{s})]^{1/3}}
\left[\int_0^{z_{s}} \frac{dz}{E(z)} \right]^{2/3}\,,
\end{equation}
with $E(z)=H(z)/H_{0}$ the normalized Hubble rate. The previous
estimator is measured from the SDSS data to be $A=0.469\pm 0.017$,
where $z_{s}=0.35$ [or $a_{s}=(1+z_{s})^{-1}\simeq 0.75$].
Therefore, the corresponding $\chi^{2}_{\rm BAO}$ function can be
written as:
\begin{equation}
\chi^{2}_{\rm BAO}({\bf p})=\frac{[A({\bf p})-0.469]^{2}}{0.017^{2}}\;.
\end{equation}

Finally, a very accurate and deep geometrical probe of dark energy is the
angular scale of the sound horizon at the last scattering surface,
as encoded in the location $l_1^{TT}$ of the first peak of the
Cosmic Microwave Background (CMB) temperature perturbation spectrum.
This probe is described by the  CMB shift parameter~\cite{Bond:1997wr,Nesseris:2006er}, defined as:
\begin{equation}\label{eq:shiftparameter}
R=\sqrt{\Omega_{m}^0}\int_{a_{ls}}^1 \frac{da}{a^2 E(a)}=\sqrt{\Omega_{m}^0}\int_{0}^{z_{ls}} \frac{dz}{E(z)}\,.
\end{equation}
The measured shift parameter according to the WMAP 7-years data~\cite{WMAP}
is $R=1.726\pm 0.018$ at the redshift of the last scattering surface: $z_{ls}=1091.36$ [or
$a_{ls}=(1+z_{ls})^{-1}\simeq 9.154\times 10^{-4}$]. In this case,
the $\chi^{2}_{\rm cmb}$ function is given by:
\begin{equation}
\chi^{2}_{\rm cmb}({\bf p})=\frac{[R({\bf
p})-1.726]^{2}}{0.018^{2}}\;.
\end{equation}
Let us note that when dealing with the CMB shift parameter we have to include both the matter and radiation terms in the total normalized matter density entering the $E(z)$ function in (\ref{eq:shiftparameter}):
\begin{equation}\label{mixture}
\OM(z)=\OM^0\,(1+z)^{3}+\OR^0\,(1+z)^{4}\,.
\end{equation}
Indeed, we have  $\ORo=(1+0.227 N_{\nu})\, \Omega_{\gamma}^0$, with $N_{\nu}$ the number of neutrino species and $\Omega_{\gamma}^0\,h^2\simeq
2.47\times 10^{-5}$. Therefore, at $z_{ls}=1091.36$ and for three light neutrino species the radiation contribution amounts to some $24\%$ of the total energy density associated to matter. For a detailed discussion of the shift parameter as a cosmological probe, see e.g. ~\cite{Elgaroy07}.

In order to place tighter constraints on the corresponding parameter
space of our model, the probes described above must be combined
through a joint likelihood analysis\footnote{Likelihoods are
normalized to their maximum values. In the present analysis we
always report $2\sigma$ uncertainties on the fitted parameters. Note
also that the total number of data points used here is
$N_\text{tot}=379$, while the associated degrees of freedom is:
$dof = N_\text{tot}-n_{\rm fit}$, where $n_{\rm fit}$ is
the model-dependent number of fitted
parameters.}, given by the product of the
individual likelihoods according to: \be\label{eq:overalllikelihood} {\cal L}_{\rm tot}({\bf p})=
{\cal L}_{\rm SNIa}\times {\cal L}_{\rm H}
\times {\cal L}_{\rm BAO} \times {\cal L}_{\rm
cmb}\;, \ee which translates in an addition for the joint $\chi^2$
function:
\be\label{eq:overalllikelihoo}
\chi^{2}_{\rm tot}({\bf p})=\chi^{2}_{\rm SNIa}+\chi^{2}_{\rm H}+
\chi^{2}_{\rm BAO}+\chi^{2}_{\rm cmb}\;.
\ee

In order to proceed with our $\chi^2$ minimization procedure, we
would like to reduce as much as possible the free parameter space by
imposing a prior, specifically we fix the value of the current mass
parameter $\Omega_m^0=\rho_m^0/\rco$.
In principle, $\Omega_m^0$ is constrained by the maximum likelihood
fit to the WMAP and SNIa data in the context of the concordance
$\Lambda$ cosmology, but in the spirit of the current work, we want to use
measures which are completely independent of the dark energy
component. An estimate of $\Omega_m^0$ without conventional priors
is not an easy task in observational cosmology. However, various
authors, using mainly large scale structure studies, have attempted
to put constraints to the $\Omega_m^0$ parameter. In a rather old
paper Plionis et al.~\cite{PLC} using the motion of the Local Group
with respect to the cosmic microwave background found
$\Omega_{m}^0\simeq 0.30$. From the analysis of the power spectrum,
Sanchez et al.~\cite{San06} obtain a value $\Omega_m^0\simeq 0.24$.
Moreover,~\cite{Fel03} and~\cite{MoTu05} analyze the peculiar
velocity field in the local Universe and obtain the values
$\Omega_m^0\simeq 0.30$ and $\simeq 0.22$ respectively. In addition,
the authors of Ref.~\cite{And05}, based on the cluster
mass-to-light ratio, claim that $\Omega_m^0$ lies in the interval
$0.15-0.26$ (see also~\cite{Sch02} for a review). Therefore, there
are strong independent indications for $0.2\lesssim
\Omega_m^0\lesssim 0.3$, and in order to compare our results with
those of the flat $\Lambda$CDM we will restrict our present analysis
to the choice $\Omega_{m}^0=0.27$. If we fix the value of
$\Omega_{m}^0$, then the corresponding vector ${\bf p}$ contains
only two free parameters namely, $(q_{0},{\dot{q}_{0}}/{H_{0}})$.
Note that we sample $q_{0} \in [-1.24,0.1]$ and
${\dot{q}_{0}}/{H_{0}} \in [-5,0.2]$ in steps of 0.001.

\section{Identifying the best relaxation $F^n_m$  models from observation}\label{sec:BestModels}

In order to select the optimal relaxation models $F^n_m$, defined in
Eq.~(\ref{eq:Fmn}), from the phenomenological point of view, we are
going to perform a likelihood analysis along the lines described in
the previous section. Specifically, we will use a two-parameter fit
of our models. Namely, for any given $F^n_m$ model with fixed $m$
and $n$, we have to look for the best fit values for $q_0$ and
$\dot{q}_0/H_{0}$.

\subsection{Numerical solution of the $F^n_m$  models}\label{sec:SolvingModel}

Remember that the parameters $q_0$ and $\dot{q}_0$ are necessary in
order to establish the initial conditions for solving the
generalized Friedmann equation (\ref{eq:Friedmann1}), in which
$\rLe$ is a complicated function of the form
$\rLe(H,q,\dot{q}/H)=\rLi+\rF(H,q,\dot{q}/H)$. Using
$\dot{H}=-H^2(q+1)$ one can see that $\rLe=\rLe(H,\dot{H},\ddot{H})$
and hence the generalized Friedmann equation, despite its innocent
appearance, becomes a third order differential equation in the scale
factor\,\footnote{Although equation (\ref{eq:Eij}) indicates that,
in principle, the field equations are of fourth order, the
self-conservation of $\rF$, see (\ref{localCCF}) (and hence also of
the effective vacuum energy $\rLe$), enables us to reduce the order
of the field equations by one unit.} $a=a(t)$. Therefore, since the
current $H_0$ is known, we need to input $q_0$ and $\dot{q}_0$ for
any given relaxation model $F^n_m$. Notice also that the initial
values of $q_0$ and $\dot{q}_0$ must be consistent with the current
value of $\OLo$, which in our case is identified with
\begin{equation}\label{OmegaCCeffnow} \Omega_{\CC{\rm eff}}^0 =
\frac{\rLi+\rF(H_0,q_0,\dot{q}_0)/H_0}{\rho_c^0}=1-\Omega^0_m\,,
\end{equation}
for flat space cosmology. For the reasons indicated in
Sect.\,\ref{sect:PhysicalScales}, we will limit ourselves to analyze
the five canonical models $F^0_1$, $F^1_1$, $F^3_2, F^{2}_{1}$ and
$F^{2}_{2}$ using the combined likelihood method described in
the previous section. We present a sample of the likelihood contours in the $(q_{0}, {{\dot q}_{0}}/{H_{0}})$ plane for the individual sets of data on SNIa, CMB shift parameter, BAO and $H(z)$ in Figs. 1-3, whereas in Fig. 4 we display the combined likelihood contours for all the five models.  A numerical summary of the statistical analysis for these models is shown in Table 1. In the next section,
we provide more details of this analysis and discuss the obtained
results.

\begin{table}[ht]
\begin{center}
\caption[]{Results of the overall likelihood function analysis, equation (\ref{eq:overalllikelihoo}). The
1\textsuperscript{st} column indicates the various $F^{n}_{m}$ models. The $\CC$CDM model is included in the first row for comparison, although in this case we present the theoretical prediction based on equations (\ref{eq:qLCDM}) and (\ref{eq:qdotLCDM}). The
2\textsuperscript{nd}, 3\textsuperscript{rd} and
4\textsuperscript{th} columns show the best fit parameters and the
reduced $\chi^{2}_{\rm tot}$.  We use the prior $\Omega_m^0=0.27$,
flat space cosmology and the (arbitrarily chosen) initial vacuum
energy such that $\rLi/ \rco = -10^{60}$ for all the models. In the
final column one can find various line types appearing
in Fig. 4  where all models are plotted together.}
\label{tab:likelihood} \tabcolsep 3.pt
\vspace{0.4cm}
\begin{tabular}{ccccc} \hline \hline
Model & $q_{0}$& ${\dot q}_{0}/H_{0}$& $\chi^{2}_{\rm tot}/377$&Symbols \\ \hline
$\Lambda$CDM & $-0.595$ & $-0.887$ \\
$F^{0}_{1}$ & $-0.506\pm 0.04$ &$-0.43\pm 0.03$ &1.188& red dashed\\
$F^{1}_{1}$ & $-0.500\pm 0.03$ &$-0.36\pm 0.07$ &1.187& magenta long-dashed \\
$F^{3}_{2}$ & $-0.551\pm 0.04$ &$-0.63\pm 0.07$ &1.186& black solid \\
$F^{2}_{1}$ & $-0.678\pm 0.04$ &$-2.27\pm 0.34$ &1.188& green dotted \\
$F^{2}_{2}$ & $ -0.520\pm 0.06 $ &$-0.480 \pm 0.05 $ & 1.187 & black dotted area
\end{tabular}
\end{center}
\end{table}

\begin{figure}
\begin{center}
\includegraphics[width=0.8\textwidth]{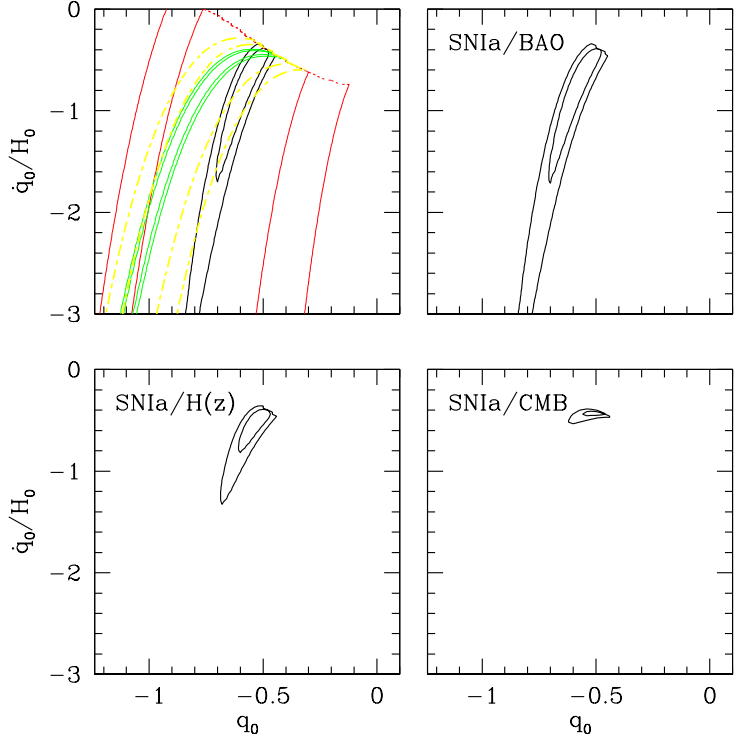}
\caption{{\it $F^{0}_{1}$ model:} Likelihood contours in the
$(q_{0}, {{\dot q}_{0}}/{H_{0}})$ plane. The contours are plotted
where $-2{\rm ln}({\cal L/L_{\rm max}})$ is equal to 2.32 and 6.16,
corresponding to $1\sigma$ and $2\sigma$ confidence level. We do not
plot the $3\sigma$ contour in order to avoid confusion. In the upper
left panel we present the likelihood contours that correspond to the
SNIa (solid lines), $H(z)$ (yellow thick dot-dashed lines),
CMB/shift parameter (green dashed lines) and BAOs observational data
(red dotted lines). The remaining panels show the statistical
results for different pairs.}\label{fig:fg01}
\end{center}
\end{figure}

\begin{figure}
\begin{center}
\includegraphics[width=0.8\textwidth]{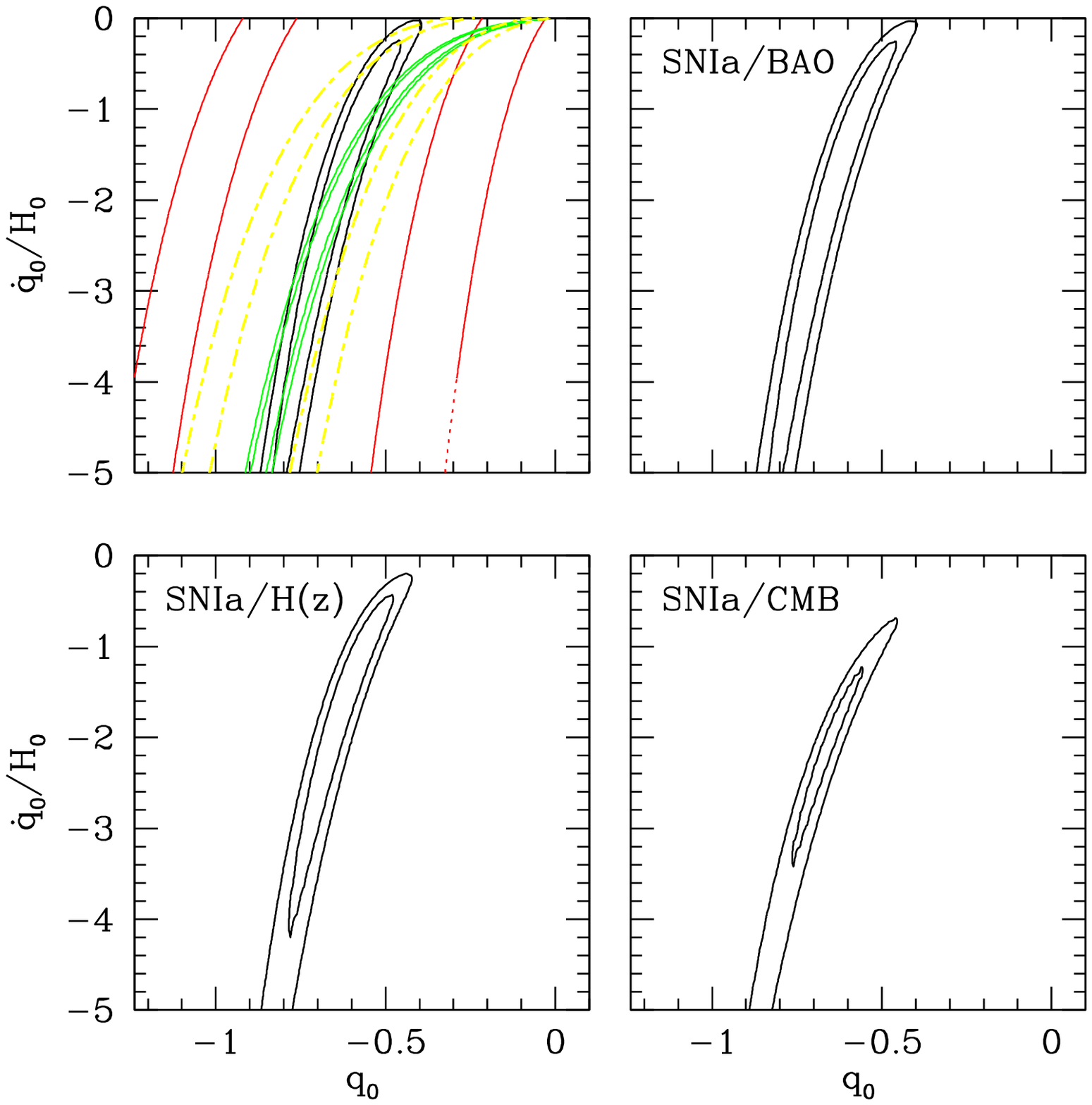}
\caption{{\it $F^{2}_{1}$ model} (or \textit{``no-scale''}
relaxation model): The Likelihood contours in the $(q_{0}, {{\dot
q}_{0}}/{H_{0}})$ plane. The different observational data are
represented by different line types (see caption of Fig.~1 for
definitions).}\label{fig:fg21}
\end{center}
\end{figure}

\begin{figure}
\begin{center}
\includegraphics[width=0.8\textwidth]{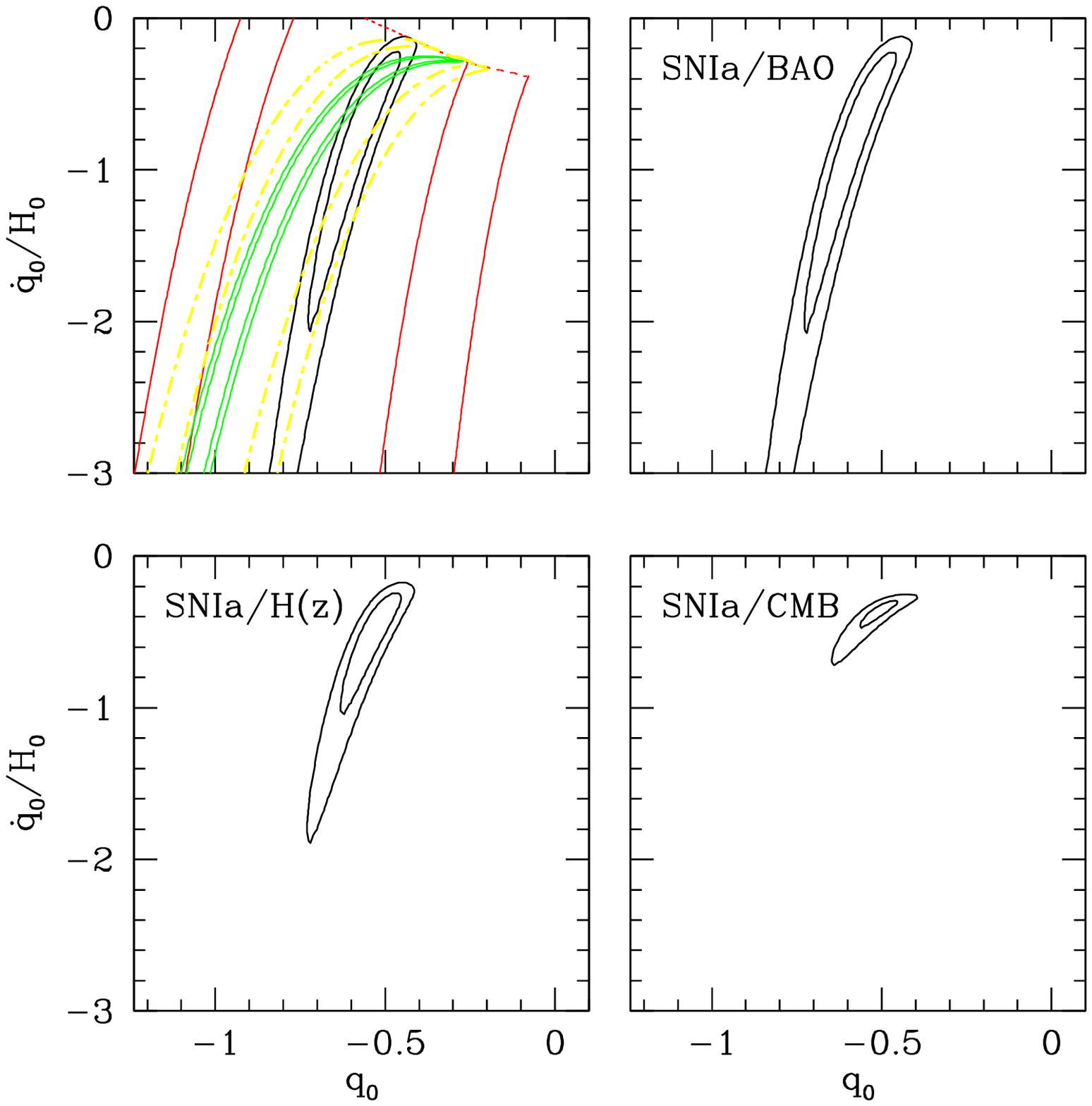}
\caption{{\it $F^{1}_{1}$ model:} The Likelihood contours in the
$(q_{0},{\dot q}_{0}/H_{0})$ plane. The different
observational data are represented by different line types (see
caption of Fig.~1 for definitions).}\label{fig:fg11}
\end{center}
\end{figure}

\subsection{The Statistical Results}\label{subsect:statistical}
In the upper left panel of Fig. \ref{fig:fg01} we present the
results of our analysis for the $F^{0}_{1}$ model in the $(q_{0},
{{\dot q}_{0}}/{H_{0}})$ plane. The individual contours for each observable are plotted only for
the 1$\sigma$ and 2$\sigma$ confidence levels in order to avoid
confusion. In particular, the SNIa-based results indicated by thin
solid lines, the $H(z)$ results by thick dot-dashed lines, the BAO
results by dotted lines and those based on the CMB shift parameter
by thin dashed lines. The remaining panels show the statistical
results for SNIa/BAO, SNIa/$H(z)$ and SNIa/CMB. Using the SNIa/BAO
data alone it is evident that the ${\dot q}_{0}/H_{0}$ parameter is
unconstrained within $2\sigma$ errors. However, within $1\sigma$
errors we can put some constraints (the best fit values are
$q_{0}\simeq -0.56$ and ${\dot q}_{0}/H_{0}\simeq -0.64$). On the
other hand, utilizing the SNIa/$H(z)$ data the best fit parameters
are partially constrained within $2\sigma$: $q_{0}\simeq -0.516$ and
${\dot q}_{0}/H_{0}\simeq -0.47$. As can be seen in the lower right
panel of Fig. \ref{fig:fg01}, the above degeneracy is broken when
using the SNIa/CMB data and practically the best fit parameters
coincide with those of joint likelihood analysis, involving all the
cosmological data. Indeed, for the $F^{0}_{1}$ model we find that
the overall likelihood function peaks at $q_{0}=-0.506 \pm 0.04$ and
${\dot q}_{0}/H_{0}\simeq -0.43\pm 0.03$ with $\chi_{\rm tot}^{2}
\simeq 447.83$ for $377$ degrees of freedom.

\begin{figure}
\begin{center}
\includegraphics[width=0.8\textwidth]{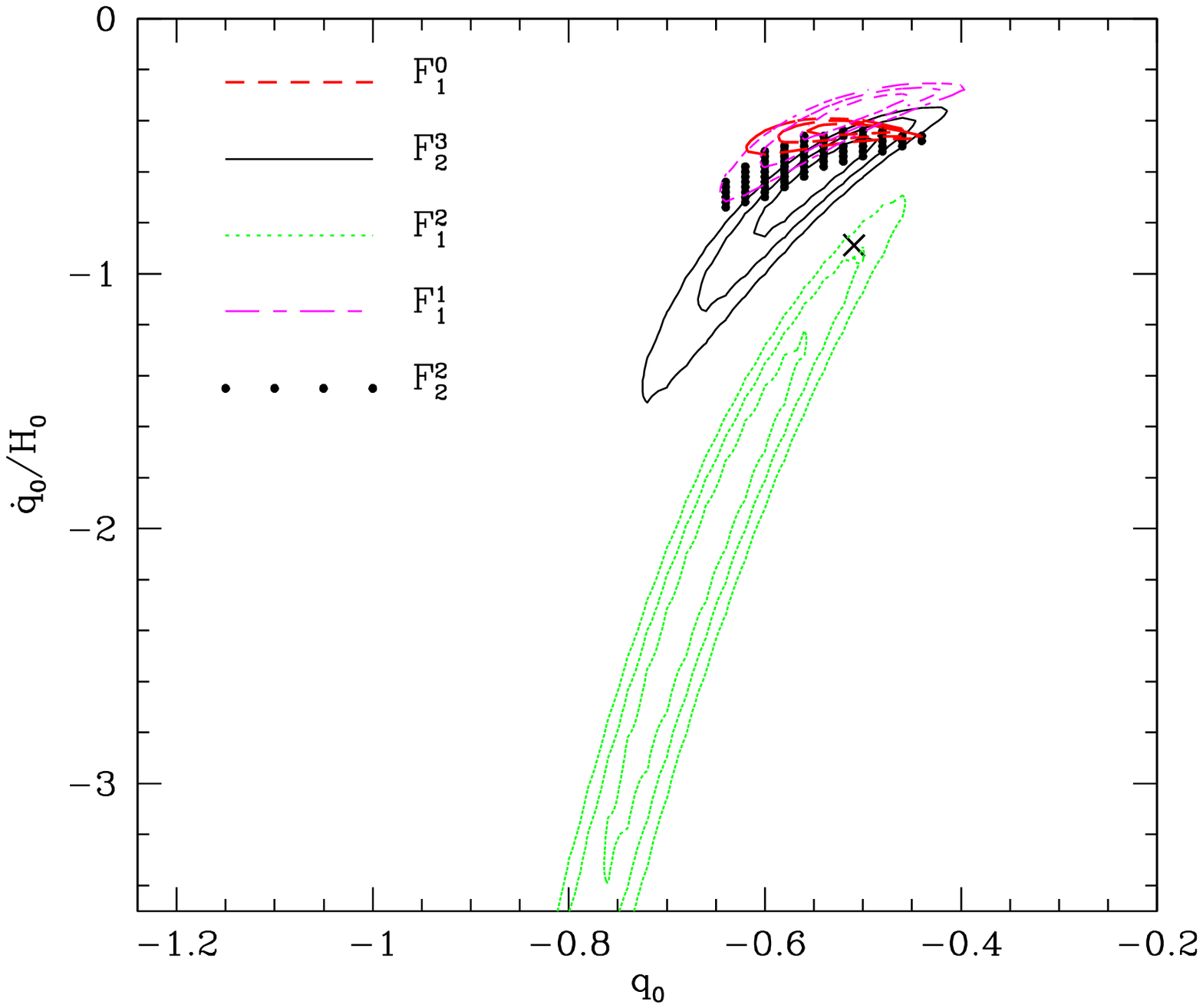}
\caption{ The overall Likelihood contours in the $(q_{0},{{\dot
q}_{0}}/{H_{0}})$ plane, computed from (\protect\ref{eq:overalllikelihood}). The contours are plotted where $-2{\rm
ln}({\cal L/L_{\rm max}})$ is equal to 2.32, 6.16 and 11.83,
corresponding to $1\sigma$, $2\sigma$ and $3\sigma$ confidence
level. The different $F^{n}_{m}$ cosmological models are represented
by the following types: $F^{0}_{1}$ red dashed line,
$F^{3}_{2}$ black solid line, $F^{2}_{1}$ green dotted line and $F^{2}_{2}$ black
dotted area (see Table I). For comparison, the cross
corresponds to the traditional
$\Lambda$CDM model.}\label{fig:fig5}
\end{center}
\end{figure}

In the case of the ``no-scale'' $F^{2}_{1}$ model we find that only
the combined SNIa/CMB data can put constraints on the free
parameters (see Fig. \ref{fig:fg21}), while the overall likelihood
analysis provides $q_{0}=-0.678 \pm 0.04$ and ${\dot
q}_{0}/H_{0}=-2.27\pm 0.34$ with $\chi_{\rm tot}^{2} \simeq 447.83$
for $377$ degrees of freedom. Concerning the $F^{1}_{1}$
model (see Fig. \ref{fig:fg11}) we
find that the corresponding statistical results are in a very good
agreement to those of the $F^{0}_{1}$ model. Indeed, the SNIa/BAO
data put some constraints within $1\sigma$ errors, $q_{0}\simeq
-0.56$ and ${\dot q}_{0}/H_{0}\simeq -0.66$ while using the
SNIa/$H(z)$ data we find $q_{0}\simeq -0.530$ and ${\dot
q}_{0}/H_{0}\simeq -0.441$. Again we observe that the joint
likelihood function (mainly due to SNIa/CMB) peaks at $q_{0}=-0.50
\pm 0.03$ and ${\dot q}_{0}/H_{0}\simeq -0.36\pm 0.07$ with
$\chi_{\rm tot}^{2} \simeq 447.5$ for $377$ degrees of freedom. As
for the $F^{3}_{2}$ model, again we find that the comparison between SNIa/BAO
does not place constraints on the free parameters while the
SNIa/$H(z)$ data put some constrains (even within $1\sigma$):
$q_{0}\simeq -0.547$ and ${\dot q}_{0}/H_{0}\simeq -0.63$. As
before, the free parameters of the model are tightly constrained by
the SNIa/CMB data. The joint likelihood function peaks at
$q_{0}=-0.551 \pm 0.04$ and ${\dot q}_{0}/H_{0}=-0.63\pm 0.07$ with
$\chi_{\rm tot}^{2} \simeq 447.12$ for $377$ degrees of freedom.
To this end for the $F^{2}_{2}$ model we find that the
SNIa/$H(z)$ comparison implies $q_{0}\simeq -0.540$ and ${\dot
q}_{0}/H_{0}\simeq -0.58$ while using the joint likelihood analysis
we find $q_{0}=-0.52\pm 0.06$ and ${\dot q}_{0}/H_{0}=-0.48\pm 0.05$ with
$\chi_{\rm tot}^{2} \simeq 447.53$ for $377$ degrees of freedom.
Although we do not present individual likelihood contours for the
$F^{3}_{2}$ and $F^{2}_{2}$ models, we can see their overall
likelihood contours in Fig.~\ref{fig:fig5}, together with those of
the other models.

The summarized analysis of all the five canonical relaxation models
is presented in Table~\ref{tab:likelihood} and in
Fig.~\ref{fig:fig5}. In this combined figure we plot the $1\sigma$, $2\sigma$, and also the $3\sigma$, overall likelihood contours for the five the models under consideration. In it we can see a compact presentation of all
our statistical results including their comparison with the
corresponding values of $(q_{0\Lambda},{\dot q}_{0\Lambda}/H_{0})$
for the concordance or traditional $\Lambda$CDM model. Let us note that the $(q_{\Lambda}(z),{\dot q}_{\Lambda}(z)/H_{0})$ pair for the
concordance model can be computed from the formulae:
\be\label{eq:qLCDM} q_{\Lambda}(z)=\frac{3}{2}\Omega_{m}(z)-1\,; \;\;\; \frac{{\dot
q}_{\Lambda}(z)}{H_{0}}=-\frac32\,E_{\Lambda}(z)(1+z)\,\frac{d\Omega_m(z)}{dz}=-\frac{9}{2}E_{\Lambda}(z)\Omega_{m}(z)\left[1-\Omega_{m}(z)\right]\,,
\ee where \be\label{eq:qdotLCDM}
\Omega_{m}(z)=\frac{\rho_m(z)}{\rc(z)}=\frac{\Omega_{m}^0(1+z)^{3}}{E_{\Lambda}^{2}(z)}\,;
\;\;\;\;\;\;\;\;\;
E_{\Lambda}(z)=\frac{H_{\CC}(z)}{H_0}=\sqrt{\Omega_{m}^0(1+z)^{3}+1-\Omega_{m}^0}
\;. \ee
Therefore, for $\Omega_{m}^0=0.27$ we obtain $(q_{0\Lambda},{\dot
q}_{0\Lambda}/H_{0})\simeq (-0.595,-0.887)$, as quoted in the first row of Table 1. Let us mention that we
have checked that using the earlier SNIa results (UNION) of Kowalski
et al.~\cite{Kowal08} does not change significantly the previously
presented constraints. Finally, let us clarify that the chosen
initial value $\rLi = -10^{60} \,\rco$ for the vacuum energy is
completely arbitrary and the results do not depend on it because the
relaxation mechanism is dynamical and hence works for any numerical choice of $\rLi$. However, in order to avoid
instabilities in the lengthy numerical analysis involved in the
solution of these models (in which the large quantity $\rLi$ almost
cancels against the dynamically generated numerical value of $\rF$)
it is convenient not to choose $\rLi$ exceedingly large, but apart
from this proviso any other arbitrarily selected value would do, as
we have checked.

\section{The expansion history/future for the $F^{n}_{m}$ models}\label{sec:ComparingLCDM}
\begin{figure}[t]
\begin{centering}
\includegraphics[width=0.85\textwidth,clip]{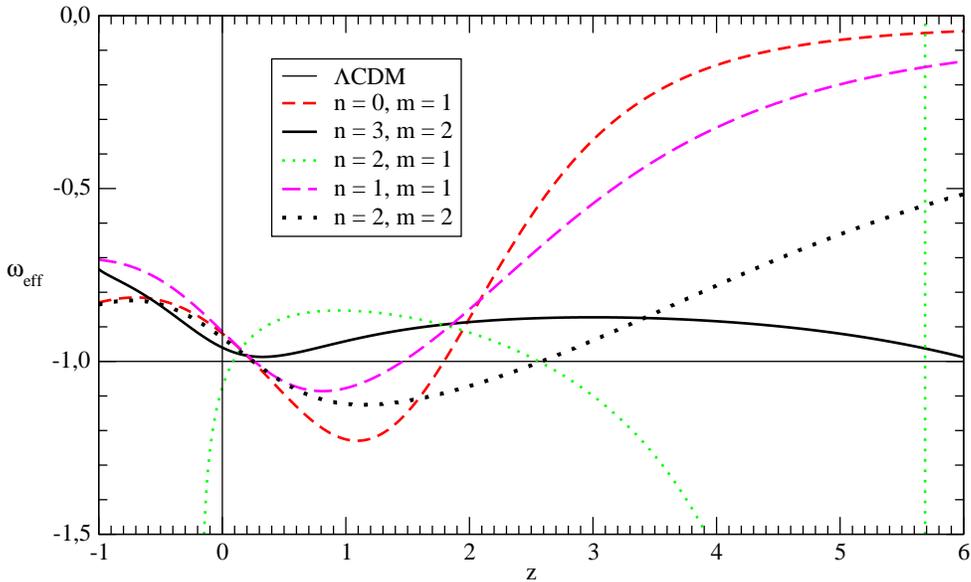}
\caption{The effective equation of state~$\omega_\text{eff}$ as a
function of the redshift~$z$ for the $F^{n}_{m}$ models with initial
conditions taken from Table~1 and with~$\Omega_m^0=0.27$. The
horizontal line $\omega_\text{eff}=-1$ corresponds to the
$\Lambda$CDM model.\label{fig:EOS-of-z}}
\end{centering}
\end{figure}

\begin{figure}[t]
\begin{centering}
\includegraphics[width=0.85\textwidth,clip]{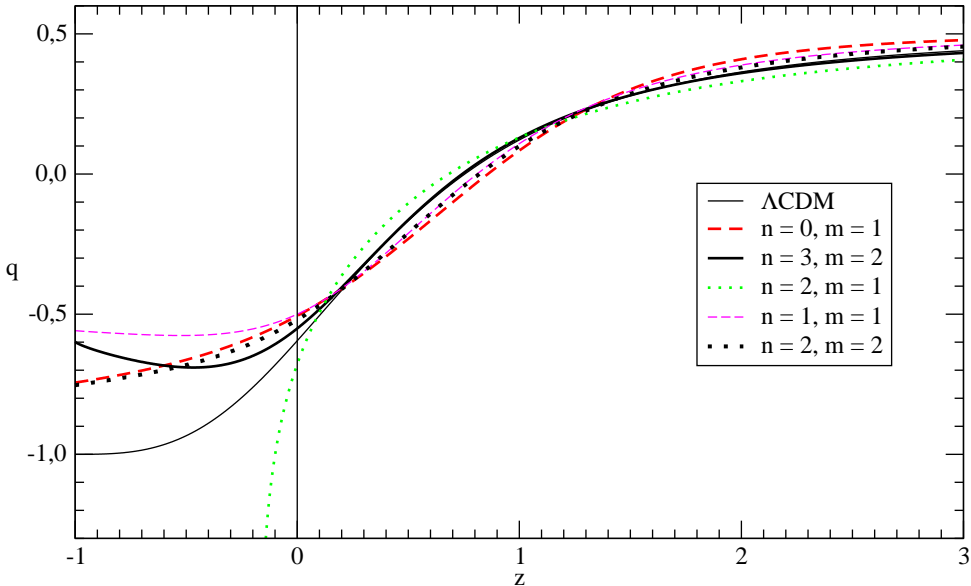}
\caption{The deceleration factor~$q(z)$ as a function of the
redshift~$z$ for the $\Lambda$CDM model with~$\Omega_m^0=0.27$ and
for the $F^{n}_{m}$ models with initial conditions taken from
Table~1. Note that for~$z>0.2$ the curves of the models~$\Lambda$CDM and~$F^3_2$ are overlapping. \label{fig:q-of-z}}
\end{centering}
\end{figure}

Since the current main cosmological quantities (scale factor, Hubble
function etc.) exhibit a complicated scaling with the redshift, the
absorption of the extra effects from the relaxation model into an
``effective dark energy'' contribution, with a non-trivial EoS of
the form~$\omega_\text{eff}(z)=\pLe(z)/\rLe(z)$ is possible with
the effective pressure~$\pLe(z)$ and energy density~$\rLe(z)$ taken
from Eqs.~(\ref{eq:rhoLambdaEff}). The corresponding EoS for the
matter component is not affected by the presence of the $F$-term, so
that the behavior of the matter epoch is the expected one. Indeed,
during the background evolution in e.g.\ the non-relativistic matter
dominated era, the deceleration parameter~$q$ changes only very
slightly to compensate the decreasing Hubble rate $H\sim t^{-1}$,
but $q$ always stays around the value~$\frac{1}{2}$. Therefore, the
universe expands like a matter dominated cosmos
despite~$|\rLi|\gg\rho_{m}$. Analogously, $q$ will vary minimally
around the value~$1$ to ensure $|\rLe|\ll|\rLi|$ during the
radiation regime. Eventually, in the asymptotic future the
relaxation of the CC is guaranteed by the smallness of~$H$ in the
function $B$ (\ref{eq:B}) at late times. The correct temporal
sequence of the three cosmic epochs follows from the different
powers of~$H$ in the function~$B$, where higher powers are more
relevant at earlier times: $\sim H^6$ for the radiation dominated
epoch, and $\sim H^4$ for the matter and vacuum dominated epochs,
although for the latter $H$ becomes smaller than in the former
because $q$ is no longer close to $1/2$. As we can see, all these
dynamical features are encoded in the structure of (\ref{eq:B}). The
precise behavior of the relevant quantities in the various epochs
becomes transparent in the various numerical examples considered in
this section (cf. Figs. \ref{fig:EOS-of-z} and \ref{fig:q-of-z}), on
which we will elaborate further below).

In general, it is well known that one can express the effective dark
energy EoS parameter in terms of the Hubble rate, $H(z)$. This
function of the cosmological redshift $z$ becomes known
(numerically) after we explicitly solve the model as indicated in
Sect.\,\ref{sec:BestModels}. In the present case the structure of
the effective EoS of the DE is quite cumbersome. First of all, from
the formulae of Sect.\,\ref{sec:CCRelax-Model} one can show that
\begin{equation}\label{eq:omegaeff}
\weff(z)=\frac{\pLe(z)}{\rLe(z)}=-1+\left[1+\wF(z)\right]\,\frac{\rF(z)}{\rLi+\rF(z)}=-1+2\,\frac{E_{\,\,
0}^{0}(z)-\frac13\,E_{\,\, i}^{i}(z)}{\rLi+2\,E_{\,\, 0}^{0}(z)}\,,
\end{equation}
where
\begin{equation}\label{eq:omegaF}
\omega_F(z)=\frac{p_F(z)}{\rF(z)}=-\frac13\,\frac{E_{\,\,
i}^{i}(z)}{E_{\,\, 0}^{0}(z)}\,.
\end{equation}

In view of equations (\ref{eq:E00}) and (\ref{eq:Eij}), the previous
expressions are complicated functions of $z$. From
(\ref{eq:omegaeff}) and (\ref{eq:omegaF}) it is patent that if we
would have $E_{\,\, 0}^{0}=(1/3)\,E_{\,\, i}^{i}$, then the EoS of
the induced DE density would be $\wF=-1$, and at the same time the
EoS of the measurable effective vacuum energy would also be
$\weff=-1$. However, as we can see from equations (\ref{eq:E00}) and
(\ref{eq:Eij}), the relation $E_{\,\, 0}^{0}=(1/3)\,E_{\,\, i}^{i}$
is only satisfied by the $F$-terms on the respective \textit{r.h.s.}
of these equations, but not by the remaining terms. Therefore, the
effective EoS of the measurable DE density is expected to have a
non-trivial behavior, and we must check if this behavior is still
sufficiently close to the $\CC$CDM result $\omega_{\CC}=-1$,
particularly at $z=0$, and also determine what is its behavior as a
function of the redshift.

In order to visualize the redshift dependence of the effective EoS
parameter, we compare in Fig.~\ref{fig:EOS-of-z} the various $F^{n}_{m}$
cosmological models indicated in Table I of Sect.\,\ref{sec:BestModels}.
One can divide the evolution of the cosmic expansion history in
different phases on the basis of the varying behavior of the
$F^{n}_{m}$ and $\Lambda$CDM models. We will investigate such
variations in terms of the deceleration parameter,
$q(z)=-1+d(\ln{H})/d(\ln{(1+z)})$, which is plotted in
Fig.~\ref{fig:q-of-z}. Below we briefly present the cosmic expansion
history of the $F^{n}_{m}$ models studied here. Note that we always
compare with the concordance $\Lambda$CDM cosmology.

\newtext{Although} we do not focus here on the details of the radiation dominated period within our
model, let us recall that it has been discussed in Sect.5.2 of Ref.\,\cite{Relax-FRG-PLB}. From the approximate effective equation of state of the radiation epoch -- see eq.(5.8) of that reference -- one can see that it behaves radiation-like, i.e. $\weff\simeq 1/3$, and smoothly connects with $\weff\simeq 0$ in the matter dominated epoch during equality. Therefore the behavior is perfectly
compatible with that of the $\CC$CDM model. This is corroborated by the numerical solution of the full field equations\,\cite{Relax-FRG-PLB}.
Finally, we point out that there is no significant effect that could alter the BBN phase of the radiation epoch because, as indicated previously, the form (\ref{eq:Fmn})-(\ref{eq:B}) insures that the dynamical relaxation mechanism automatically reduces the big initial CC to a tiny value at all epochs, starting from the radiation epoch, going through the matter epoch until our present epoch. The density of vacuum energy during BBN therefore was completely subdominant and could not have any measurable effect on
the standard light element abundances.

\newtext{Prior} to the radiation epoch we have the inflationary epoch. The latter should not be affected by the relaxation mechanism. This can be
seen once more from the general structure of the functional (\ref{eq:Fmn})-(\ref{eq:B}). In the inflationary epoch, $H$ is very large and $q$ is near $-1$. Therefore the behavior of $F$ in equation (\ref{eq:Fmn}) is
of the form $F\sim 1/H^{6m-2n}$, which
after applying the natural relaxation condition (\ref{eq:NatRelax}) implies
it is severely suppressed as $|F|\leq 1/H^{2m}$, at least. Thus, the universe is then completely controlled by the physics of the inflationary epoch, whatever it be, without receiving any interference from our relaxation mechanism. This mechanism only starts working when the universe leaves the
inflationary epoch and enters the radiation dominated one, since
then $q\to +1$ and this triggers the first large contribution from
$F$ that compensates for the value of the huge CC left over at the end of the
inflationary period. Although we do not attempt to describe here how the
inflationary epoch transited into the radiation dominated one,
in the original formulation of the relaxation model [4] it was suggested
that the functional $F$ in equation (\ref{eq:Fmn}) could contain an additive polynomial $A(R)$ of the Ricci scalar. In its simplest non-trivial form it would just entail a term proportional to $R^2$, which would not alter the relaxation mechanism in the radiation epoch. The advantage of this addition is that it could
connect this relaxation model with a Starobinsky's type of
mechanism for inflation\,\cite{Starobinsky80} and subsequent modifications thereof, see e.g. \cite{Fossil07} and references therein.

\subsection{Cosmic acceleration and effective EoS analysis}\label{sec:qweff}

In the following, we remark some features that can be observed from
the behavior of the effective EoS and deceleration parameter of the
various relaxation models under consideration in
Figs.~\ref{fig:EOS-of-z} and~\ref{fig:q-of-z}, respectively:

\begin{itemize}

\item {\bf $F^{0}_{1}$ model:} For $z\ge 1$ the deceleration parameters
(both for $F^{0}_{1}$ and $\Lambda$CDM) are positive with
$q(z)\gtrsim q_{\Lambda}(z)$, which means that the
    cosmic expansion in the $F^{0}_{1}$ model is more rapidly
    ``decelerating'' than in the $\Lambda$CDM case.
At $z\sim 0.8$ both models are starting to
    accelerate. Between
$0.2\lesssim z \lesssim 0.8$ the deceleration parameters are
    both negative with $q(z)<q_{\Lambda}(z)$.
Now we will focus on the evolution of the effective EoS parameter.
In particular, we find that at early enough times $z\ge 3$ the
effective EoS parameter~$\omega_\text{eff}(z)$ of the $F^{0}_{1}$
model approaches zero\,\footnote{This is no surprise, it is actually
a common feature expected for general $\Lambda$XCDM models (as the
relaxation models indeed are), see \cite{LXCDM} for details.}, while
we always have $\omega_{\Lambda}(z) = -1$ for the $\Lambda$CDM
model. At $z\simeq 1.8$ the $F^{0}_{1}$ effective EoS parameter
crosses the phantom divide line ($\omega_\text{eff}=-1$) and it
stays there for some time ($0.2\lesssim z \lesssim 1.8$). Close to
the present epoch, the effective EoS parameter crosses the phantom
divide again and it behaves quintessence-like at present, where
$\omega_\text{eff}(0)\simeq -0.92$. In the future, it
sustains this quintessence-like behavior and tends to
$\omega_\text{eff}^{\infty}\simeq -0.83$ .

\item {\bf $F^{1}_{1}$ model:} this case behaves qualitatively very similarly to
the $F^{0}_{1}$ model with a transient $\omega_\text{eff}<-1$ phase
in the past and quintessence like behavior in the asymptotic future.
It also behaves quintessence-like at $z=0$, where it takes
essentially the same value of $\omega_\text{eff}(0)\simeq -0.91$ as in the
previous model. The asymptotic EoS behavior is
$\omega_\text{eff}^{\infty}\simeq -0.6$.

\item {\bf $F^{3}_{2}$ model:} for $0\le z \le 3$ the
deceleration parameter practically coincides with that of the
concordance $\Lambda$ cosmology, which means that the cosmic
expansion in the $F^{3}_{2}$ model mimicks that of the $\Lambda$CDM
case. Also, the effective EoS is not far from~$-1$ in this region,
but it stays above the phantom divide line. Specifically, the
current value is $\omega_\text{eff}(0)\simeq -0.96$.  For
$z<0$ (future) the deceleration parameters are both negative with
$q(z)>q_{\Lambda}(z)$, which means that the cosmic expansion in the
$\Lambda$CDM model is more rapidly ``accelerating'' than in the
$F^{3}_{2}$ case. The latter approaches a quintessence like future
with~$\omega_\text{eff}^{\infty}\simeq -0.73$.

\item {\bf $F^{2}_{1}$ model} (or ``no-scale'' model): for $1\le z \le 3$ the
deceleration parameter remains close to that of the concordance
$\Lambda$ cosmology. Then at $z\sim 0.7$ both models are starting to
accelerate and between $0\lesssim z \lesssim 0.7$ the deceleration
parameters are both negative with $q(z)>q_{\Lambda}(z)$. Near
the present epoch, the effective EoS parameter $\omega_\text{eff}
\sim -1$, in particular the current value is
$\omega_\text{eff}(0)\simeq -1.08$. Interestingly, in the
future the $F^{2}_{1}$ leads to an apparent singularity
($\omega_\text{eff}(z) \ll -1$) caused by the pole that
$\omega_\text{eff}$ has at $z\sim 5.7$ due to~$\rLe$ changing its
sign. There is, however, no physical divergence related to it
because all energy densities remain finite. The dramatic change of
qualitative behavior of this model with respect to the others is
nevertheless remarkable. In particular, the fact that $F^{2}_{1}$
behaves mildly phantom-like at present could explain the persistent
tilt of the EoS data pointing slightly below $-1$ at $z=0$, if
eventually confirmed by the observations.

\item {\bf $F^{2}_{2}$ model:} close to the present time and in the future this model performs very similarly to the $F^0_1$ model with $\omega_\text{eff}(0)\simeq -0.93$. However, the transient $\omega_\text{eff}<-1$ phase between $z\approx 0.3$ and $z\approx 2.5$ started earlier in the past, roughly at the same redshift where the $F^2_1$ model switched from the phantom- to the quintessence-like EoS.

\end{itemize}

\subsection{Phase space analysis for the ``no-scale'' models}\label{sec:PhaseSapce}

\begin{figure}
\begin{centering}
\includegraphics[width=0.91\textwidth]{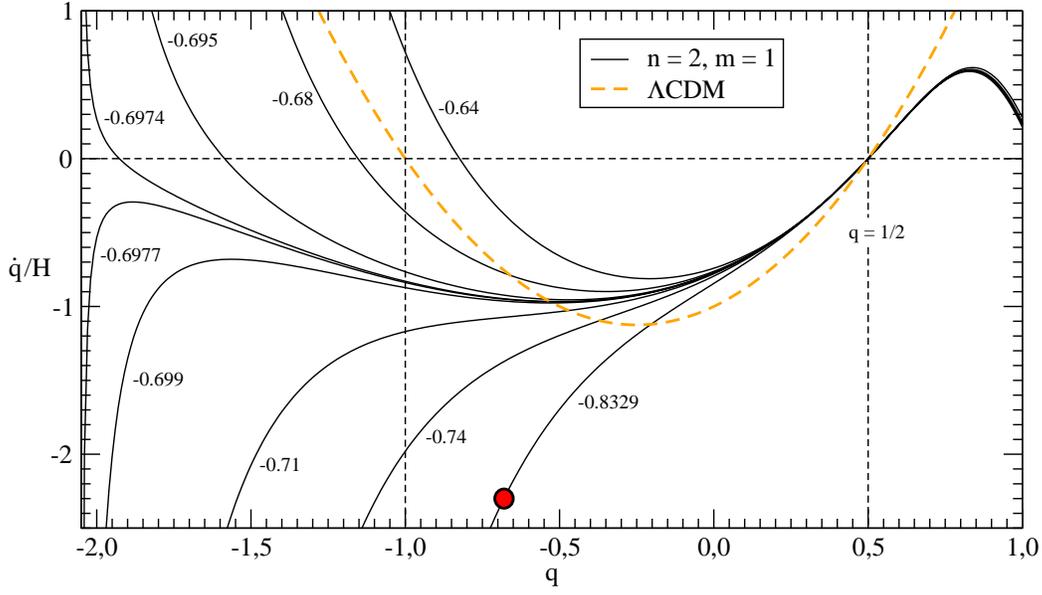}
\par\end{centering}
\caption{\label{fig:qd-q-phasespace} Phase space diagram for the
case~$n=2$, $m=1$. We plot $\dot{q}/H$ as a function of the
deceleration~$q$ for different values of
$\rho_{\Lambda}^{i}/(2\beta)$ (attached to the curves). The matter
cosmos with~$q=\frac{1}{2}$ (right vertical dashed line) is an
attractor in the past. The red blob denotes the best-fit point
($q_{0}=-0.68$, $\dot{q}_{0}=-2.3$) found in the statistical
analysis of Sect.~3. Moreover, we show~$\dot{q}/H$ (dashed orange
curve) for the $\Lambda$CDM model without radiation, which has also
a stable fixed point at~$q=-1$ corresponding to de~Sitter space-time
(left vertical dashed line). In this plot $q$ is
decreasing/increasing with time in the lower/upper part divided by
the horizontal dashed line.}
\end{figure}

The ``no-scale'' models  $n=2m$, like the $F^2_1$ discussed in the
previous section, deserve a closer attention. From the peculiar
structure of (\ref{rFmeqn}), we see that $\rF$ is a function of only
two arguments: $\rF=\rF(q,\dot{q}/H)$. This fact enables us to
analyze the $F^{2m}_m$ models in a $(q,\dot{q}/H)$ phase space
diagram. This is generally not the case for the other models, for
which $\rF$ is in general a function of three (not just two)
independent arguments $\rF=\rF(H, q,\dot{q}/H)$, see equation
(\ref{rFgeneral}) and also Appendix~\ref{sect:rFmn}. If we focus
once more on the canonical case $F^2_1$, we explicitly obtain
\begin{equation}\label{eq:rF12}
\rho_{F}(q,\dot{q}/H)=\frac{6\beta}{b^{2}}\left[(q^{4}-4q^{3}+9q^{2}-5q-1)-(2q^{3}-6q+5)\frac{\dot{q}}{Hb}\right]\,,
\end{equation}
with $\beta=M_X^4$ and $b$ given by (\ref{eq:b}). If we compare with
(\ref{rFgeneral}), here we have
$f(q)=6(q^{4}-4q^{3}+9q^{2}-5q-1)/{b^{2}}$ and
$g(q)=-6(2q^{3}-6q+5)/b^3$, and from the value of $q_0$ taken by
this model (see Table 1) we can see that $f(q_0)$ and $g(q_0)$ are
of order 1 for that matter. The corresponding phase space diagram is
depicted in Fig.\,\ref{fig:qd-q-phasespace}. The background
evolution follows mainly from $\rho_{F}+\rho_{\Lambda}^{i}=0$, which
yields~$\dot{q}/H$ as a function of only~$q$, see
Fig.~\ref{fig:qd-q-phasespace}.

When~$\dot{q}/H$ is very small, the
deceleration does not change much. This happens
around~$q\approx\frac{1}{2}$ corresponding to the matter era. For
smaller values of $q\lesssim\frac{1}{2}$, the time
derivative~$\dot{q}$ is negative and $q$ will decrease with time.
Finally, $\dot{q}$ will vanish again in another fixed point, which
can be quintessence-like ($q>-1$) or phantom-like ($q<-1$), or
$\dot{q}/H\rightarrow-\infty$ diverges at around~$q\approx-2.05$.
The type of final state depends on the value of
$\rho_{\Lambda}^{i}/(2\beta)$ (the values are attached to the curves
in Fig.~\ref{fig:qd-q-phasespace}) or equivalently on the current
values of~$q_{0}$ and~$\dot{q}_{0}$. The smallness of the late-time
Hubble rate~$H_{0}$ results from the standard evolution
of~$H\approx\frac{2}{3}t^{-1}$ during the matter epoch, which
corresponds to the unstable fixed point ($q=\frac{1}{2}$,
$\dot{q}=0$).

\section{Conclusions}\label{sec:Conclusions}

In this paper, we have confronted the class of modified gravity
models of the form $F^n_m(R,\GB)$ -- see equation\,(\ref{eq:Fmn}) --
with the main observational data. We have shown that the set of all
models of this kind with $n\leq 2m$ are possible candidates capable
to implement the dynamical relaxation of a large cosmological term,
or vacuum energy density, irrespective of its value and size --
typically $\rLi\sim M_X^4$ (with $M_X\geq 10^{16}$ GeV). We have
adopted the point of view that such a huge vacuum energy is an
integral part of the energy budget of the universe, which was
unavoidably deposited in it by the quantum fluctuations and the
important phase transitions of the early times. All the usual
modified gravity models we know of, and in fact
most DE models of all kinds proposed in the literature, implicitly
assume that this gigantic energy is just canceled by some extremely
fine tuned counterterm such that the measured value is the tiny
number $\rLo\sim 10^{-47}$ GeV$^4$ (in particle physics units),
otherwise the cosmos could not evolve in the observed standard
manner. Usually the measured value is then linked to some late-time effective gravity modification or to the residual value of some scalar field potential. However, the preposterous cancelation that was implicitly assumed to get rid of the huge initial vacuum energy density is generally recognized by theoretical physicists as completely unnatural and
unacceptable.

Quite in contrast, the relaxation mechanism that we
have put to the observational test in this paper is able to
counterbalance the large vacuum energy (whatever its value and size) in a totally dynamical way (hence without any sort of fine tuning) thanks to an effective action which is modified by the presence of the $F^n_m(R,\GB)$
terms. By carefully comparing these relaxation models with the most
recent observational data on SNIa, CMB, BAO and high redshift
$H(z)$, we have shown in particular that the simplest candidates in the list, namely
$F^0_1$, $F^1_1$, $F^3_2$, $F^2_1$ and $F^{2}_2$, can provide
a background cosmic evolution very close to the
concordance $\CC$CDM model. In other words, these $F^n_m$ models
yield a fairly standard-like cosmological evolution going through
the normal radiation dominated, matter dominated, and CC dominated
epochs, they also exhibit an effective EoS behavior very near to $-1$ around $z=0$,  and finally (and this is of course the main point to be stressed here) they do all this without any need of enforcing fine tuning in order to get rid of the huge vacuum energy injected in the cosmos during the early
stages of its evolution. In particular, the ``no-scale'' relaxation
models, i.e.\ the entire class $F^{2m}_m$, can provide a natural
relaxation of the vacuum energy without introducing any other energy
scale except that of the vacuum energy itself. The other $F^n_m$
models, with $n\neq 2m$, involve a mass scale which, for the
canonical candidates mentioned above, is of the order of a typical
Particle Physics scale in the SM of electroweak and strong
interactions. Furthermore, in contrast to quintessence-like models,
these relaxation models do not involve extremely tiny mass scales.
Actually, the typical scales that are required are fixed in order
of magnitude only, and therefore there is no need at all of
exerting fine-tuning in any of the $F^n_m$ models so as to
dynamically reduce the huge initial value of the vacuum energy
density $\rLi\sim M_X^4$ into the very small one, $\rLo\sim
10^{-47}$ GeV$^4$, measured in our current universe. The class of models studied here can be thought of as a prototype for eventually solving the ``old CC problem'' without departing significantly from the expansion history of the concordance model. However, more realistic models are of course needed in order to provide a true solution to the cosmological constant problem from the point of view of fundamental physics.

\vspace{0.0cm}

\acknowledgments This work has been supported in part by MEC and
FEDER under project FPA2010-20807, by the Spanish Consolider-Ingenio
2010 program CPAN CSD2007-00042 and by DIUE/CUR Generalitat de
Catalunya under project 2009SGR502. SB wishes to thank the Dept.\
ECM of the Univ.\ de Barcelona for the hospitality, and the
financial support from the Spanish Ministry of Education, within the
program of Estancias de Profesores e Investigadores Extranjeros en
Centros Espa\~noles (SAB2010-0118).

\appendix

\section{Computing $\rF$ for the canonical models: $F^0_1, F^1_1, F^2_1, F^3_2, F^2_2$}\label{sect:rFmn}

Here we quote the explicit expressions for~$\rF=2E_{\,\,0}^{0}$ for
our favorite $F^n_m$ models in Table 1 of
Sect.~\ref{subsect:statistical}, which are obtained from explicit
computation of (\ref{eq:E00}). To avoid too lengthy expressions we
set~$y=0$ in equation (\ref{eq:B}) as it is not important for the
matter dominated epoch or any time after it, and in this way they
all take the general form $\rF(H,q,\dot{q}/H)$ given in
(\ref{rFgeneral}), with the notation (\ref{eq:b}). The explicit
results read as follows:
\begin{enumerate}
\item Case $n=0$, $m=1$, with $|\beta|=\M^{8}$:
\begin{equation}
{\rho_{F}}=\frac{\beta}{2b^{2}H^4}\left[(5q^{2}-3q-5)-(4q^{2}-10q+7)\frac{\dot{q}}{H
b}\right]\,;
\end{equation}

\item Case $n=1$, $m=1$, with $|\beta|=\M^{6}$:
\begin{equation}
{\rho_{F}}=\frac{\beta}{b^{2}H^2}\left[-\frac32(4q^{3}-7q^{2}+2q+4)+(4q^{3}-12q^{2}+18q-11)\frac{\dot{q}}{H
b}\right]\,;
\end{equation}

\item Case $n=2$, $m=1$, with $|\beta|=\M^{4}$ (see Sect.~\ref{sect:moreFnm}  for a detailed discussion of
this special case):
\begin{equation}
{\rho_{F}}=\frac{6\beta}{b^{2}}\left[(q^{4}-4q^{3}+9q^{2}-5q-1)-(2q^{3}-6q+5)\frac{\dot{q}}{Hb}\right]\,.
\end{equation}

\item Case $n=3$, $m=2$, with $|\beta|=\M^{6}$:
\begin{eqnarray}
{\rho_{F}}=\frac{3(q-1)\beta}{b^{3}H^2}\Big[-3(q&-&1)(4q^{3}-6q^{2}+5q+6)\nonumber\\
&+&\left.(4q^{4}-12q^{3}+28q^{2}-36q+19)\frac{\dot{q}}{Hb}\right]\,;
\end{eqnarray}

\item Case $n=2$, $m=2$, with $|\beta|=\M^{8}$:
\begin{eqnarray}
\rho_F = \frac{\beta}{8\,b^3H^4}\Big[(30 q^4 -73 q^3 &+&44 q^2
+37 q -38)\nonumber\\
&-& 2 (12 q^4 -48 q^3 +88 q^2 -82 q +31)\,\frac{\dot{q}}{Hb} \Big]
\,.
\end{eqnarray}

\end{enumerate}
In all cases $\M$ is a mass scale associated to the relaxation
mechanism, which depends on the particular model, see equation
(\ref{eq:Mpower}). The numerical value of that scale (fixed in order
of magnitude only) is determined by the value of $\rLi$. The models
listed above have been chosen because they all satisfy the ``natural
relaxation condition'' (\ref{eq:NatRelax}), and for all of them $\M$
takes a characteristic value which ranges from a mass scale of the
SM of Particle Physics to the GUT scale associated the big initial
value of the vacuum energy, $\rLi\sim M_X^4$. See sections
~\ref{sect:PhysicalScales} and \ref{sect:moreFnm}  for details.

\newcommand{\JHEP}[3]{ {JHEP} {#1} (#2)  {#3}}
\newcommand{\NPB}[3]{{ Nucl. Phys. } {\bf B#1} (#2)  {#3}}
\newcommand{\NPPS}[3]{{ Nucl. Phys. Proc. Supp. } {\bf #1} (#2)  {#3}}
\newcommand{\PRD}[3]{{ Phys. Rev. } {\bf D#1} (#2)   {#3}}
\newcommand{\PLB}[3]{{ Phys. Lett. } {\bf B#1} (#2)  {#3}}
\newcommand{\EPJ}[3]{{ Eur. Phys. J } {\bf C#1} (#2)  {#3}}
\newcommand{\PR}[3]{{ Phys. Rep. } {\bf #1} (#2)  {#3}}
\newcommand{\RMP}[3]{{ Rev. Mod. Phys. } {\bf #1} (#2)  {#3}}
\newcommand{\IJMP}[3]{{ Int. J. of Mod. Phys. } {\bf #1} (#2)  {#3}}
\newcommand{\PRL}[3]{{ Phys. Rev. Lett. } {\bf #1} (#2) {#3}}
\newcommand{\ZFP}[3]{{ Zeitsch. f. Physik } {\bf C#1} (#2)  {#3}}
\newcommand{\MPLA}[3]{{ Mod. Phys. Lett. } {\bf A#1} (#2) {#3}}
\newcommand{\CQG}[3]{{ Class. Quant. Grav. } {\bf #1} (#2) {#3}}
\newcommand{\JCAP}[3]{{ JCAP} {\bf#1} (#2)  {#3}}
\newcommand{\APJ}[3]{{ Astrophys. J. } {\bf #1} (#2)  {#3}}
\newcommand{\AMJ}[3]{{ Astronom. J. } {\bf #1} (#2)  {#3}}
\newcommand{\APP}[3]{{ Astropart. Phys. } {\bf #1} (#2)  {#3}}
\newcommand{\AAP}[3]{{ Astron. Astrophys. } {\bf #1} (#2)  {#3}}
\newcommand{\MNRAS}[3]{{ Mon. Not. Roy. Astron. Soc.} {\bf #1} (#2)  {#3}}
\newcommand{\JPA}[3]{{ J. Phys. A: Math. Theor.} {\bf #1} (#2)  {#3}}
\newcommand{\ProgS}[3]{{ Prog. Theor. Phys. Supp.} {\bf #1} (#2)  {#3}}
\newcommand{\APJS}[3]{{ Astrophys. J. Supl.} {\bf #1} (#2)  {#3}}

\newcommand{\Prog}[3]{{ Prog. Theor. Phys.} {\bf #1}  (#2) {#3}}
\newcommand{\IJMPA}[3]{{ Int. J. of Mod. Phys. A} {\bf #1}  {(#2)} {#3}}
\newcommand{\IJMPD}[3]{{ Int. J. of Mod. Phys. D} {\bf #1}  {(#2)} {#3}}
\newcommand{\GRG}[3]{{ Gen. Rel. Grav.} {\bf #1}  {(#2)} {#3}}



\end{document}